# 3D Bayesian Variational Surface Wave Tomography and Application to the Southwest China

**Wenda Yang** [1], **Xin Zhang** [1*]

[1] State Key Laboratory of Deep Earth Exploration and Imaging, School of Engineering and Technology, China University of Geosciences, Beijing 100083, China

* Corresponding author: Xin Zhang. E-mail: xzhang@cugb.edu.cn

**Abstract**

Seismic surface wave tomography uses surface wave information to obtain velocity structures in the subsurface. Due to data noise and nonlinearity of the problem, surface wave tomography often has non-unique solutions. It is therefore required to quantify uncertainty of the results in order to better interpret the resulting images. Bayesian inference is the most widely-used method for this purpose. However, the commonly-used Monte Carlo methods require huge computational cost and remains intractable in high-dimensional problems. Variational inference uses optimization to solve Bayesian inverse problems, and therefore can be more efficient in the case of large datasets and high-dimensional parameter spaces. Variational inference has been applied to 2-D surface wave tomographic problems. In this study, we extend the method to 3-D surface wave tomography by directly inverting for 3-D spatial structures from frequency-dependent travel time measurements. Specifically, we apply three variational methods, mean-field automatic differential variational inference (mean-field ADVI), physically structured variational inference (PSVI) and stochastic Stein varational gradient descent (sSVGD) to surface wave tomographic problems using both synthetic data and real data in the Southwest China. The results show that all methods can provide accurate velocity estimates, while sSVGD produces more reasonable uncertainty estimates than mean-field ADVI and PSVI because of Gaussian assumption used in the later methods. In the real data case, the variational methods provide more detailed velocity structures than those obtained using traditional methods, along with reliable uncertainty estimates. We therefore conclude that variational surface wave tomograph can be applied fruitfully to many realistic problems.

## 1. Introduction

Understanding property variations of the Earth's crust is important to geoscientists, because these variations can provide critical insights into tectonic processes, seismic hazard assessment, natural resource exploration, and environmental monitoring. Seismic surface wave tomography is one popular method used to study these variations. Surface waves are low-frequency seismic waves that travel along the Earth's surface, and are dispersive, meaning that their velocity depends on the frequency of the wave (Aki & Richard 2002). By analyzing this dispersion characteristics of surface waves, one can infer velocity structure beneath the surface. Seismic surface wave tomography has been applied in various geological settings, from global-scale studies of the Earth's mantle to local investigations of the crust (Trampert & Woodhouse 1995; Shapiro & Ritzwoller 2002; Meier *et al*. 2007a,b; Zielhuis & Nolet 1994; Curtis *et al*. 1998; Simons *et al*. 2002).

Seismic surface wave tomography can be conducted in either a two-step strategy or a one-step strategy. In the two-step strategy, one first constructs a series of 2-D phase or group velocity maps for different periods tomographically at each geographic point of interest using arrival times measured at each period; the 1-D dispersion curve at each geographic point is then inverted to obtain a 1-D shear velocity model beneath this point. Those 1-D shear velocity structures are finally interpolated to construct a 3-D model (Nakanishi & Anderson 1983; Trampert & Woodhouse 1995). Since the two steps are conducted separately and sequentially, the solution to the 1-D depth inversion cannot interact directly with the 2-D phase or group velocity tomography step. In addition, in the second step the 1-D dispersion inversions are conducted separately, and consequently horizontal spatial correlation information can be lost in the obtained 3-D model. As a result, the results obtained using this strategy may be biased (Zhang *et al.* 2018). By contrast, in the one-step strategy one constructs a 3-D shear velocity model directly from frequency-dependent travel time measurements, which can naturally impose spatial correlations in the model and produce more accurate results (Fang *et al.* 2015; Zhang *et al.* 2018). In either strategy, surface wave tomography involves solving one or a set of inverse problems.

Surface wave inverse problems are traditionally solved using optimization, in which one seeks an optimal solution by iteratively perturbing a set of starting model parameters to minimize the misfit between observed data and data predicted by the known physics from the model (Trampert & Woodhouse 1995; Ritzwoller *et al*. 2002). Since most tomographic problems are non-linear and under-determined, regularization is often imposed to incorporate additional constraints on model parameters. (Aki & Lee 1976; Dziewonski & Woodhouse 1987; Iyer & Hirahara 1993; Tarantola 2005). However, the regularization is often chosen by ad hoc means (often trial and error), and valuable information in the data can be concealed by the regularization (Zhdanov 2002). In addition, since optimization methods typically provide a single optimal solution, it is difficult to quantify uncertainty in these methods. This limitation is particularly problematic in fields like geophysics and engineering, where understanding the range of possible solutions and their likelihood is crucial for robust decision-making.

To address these challenges, Bayesian inference methods have been introduced to geophysics. Unlike traditional optimization methods, Bayesian inference treats the unknown parameters as random variables and uses Bayes' theorem to update a prior distribution with new information contained in the data to obtain a so-called posterior probability density functions (pdfs), which describes uncertainty in the solution. A common way to solve Bayesian inference problems is Markov chain Monte Carlo (McMC) sampling. The method generates a set (or chain) of samples from the posterior pdf by taking a structured random walk through a parameter space (Geyer 2011; Bishop & Nasrabadi 2006; Andrieu & Thoms 2008); thereafter these samples can be used to estimate useful information about that pdf (mean, standard deviation, *etc*.). The Metropolis-Hastings (MH) algorithm (Chib & Greenberg 1995) is one such method, which has been widely applied in geophysics to solve non-linear inverse problems (Mosegaard & Tarantola 1995; Sambrige & Mosegaard 2002), such as seismic travel time tomography (Debski 2004; Zhang *et al.* 2020 a, b), surface wave dispersion inversion (Molnar *et al.* 2010; Gosselin *et al.* 2017), and full waveform inversion (Kotsi *et al.* 2020; Fu & Innanen 2021; Zhu 2018). The method has further been extended to trans-dimensional inversion, called the reversible jump McMC (rj-McMC) method (Green 1995), which allows for changes in the dimensionality of the model space, and consequently the parameterization itself can be adapted to the data. The rj-McMC algorithm has been widely used to solve 1-D dispersion inversion (Bodin *et al.* 2012; Young *et al.* 2013; Galetti *et al.* 2017; Zhao *et al.* 2022; Gao & Lekić 2018) and 2-D phase or group velocity tomographic problems (Bodin & Sambridge 2009; Galetti *et al.* 2015, 2017; Burdick & Lekic´ 2017; Zhao *et al.* 2022; Kästle *et al.* 2024). However, due to its random-walk behavior, the method becomes inefficient in high dimensional space. As a result, its application in one-step 3-D surface wave tomography has been limited (Zhang *et al.* 2018, 2019; Rahimi *et al.* 2024). To improve efficiency of McMC sampling, more advanced methods have been introduced to geophysics, such as Hamiltonian Monte Carlo (Fichtner *et al.* 2019; Gebraad *et al.* 2020; Aleardi *et al.* 2020), stochastic Newton McMC (Martin *et al.* 2012; Zhao & Sen 2021; Huang *et al.* 2020), Langevin Monte Carlo (Durmus *et al.* 2019; Izzatullah *et al.* 2021), and Gaussian mixture McMC (Tilmann *et al.* 2020; Zhang *et al.* 2021). Nevertheless, these methods still demand enormous computational resources, and therefore cannot be applied to one-step 3-D surface wave tomography.

Variational inference addresses Bayesian inference problems by approximating the posterior distribution within a specified family of pdfs. This is achieved by minimizing a measure of the difference between the posterior pdf and the approximating pdf, for example, the Kullback-Leibler (KL) divergence (Kullback & Leibler 1951). The method therefore employs an optimization-based approach rather than random sampling, and can potentially offer greater computational efficiency and improved scalability for high-dimensional problems compared to McMC methods (Blei & Jordan 2006; Blei *et al.* 2017).

In variational inference, the choice of the approximating family of pdfs is important as it determines the accuracy of the results and the difficulty of the optimization problem. Different families can lead to different inversion results (Zhang *et al.* 2021). A common choice is to use a mean-field approximation family (Bishop & Nasrabadi 2006; Blei *et al.* 2017), which decomposes the joint probability distribution into a product of mutually independent marginal distributions (Jordan *et al.* 1999). In Geophysics the method has been used to invert for the spatial distribution of geological facies given seismic data (Nawaz & Curtis 2018, 2019; Nawaz *et al.* 2020). Nevertheless, the method neglects correlation information between parameters, and usually involves tedious derivations and bespoke implementations for each type of problem which restricts their applicability (Bishop & Nasrabadi 2006;

Blei *et al.* 2017; Nawaz & Curtis 2018, 2019). To make variational methods easier to use, Kucukelbir et al. (2017) proposed an automatic differential variational inference (ADVI) method based on a Gaussian variational family, which can easily be applied to many Bayesian inference problems. For example, the method has been used to solve seismic travel time tomography (Zhang & Curtis 2020a) and earthquake slip inversion problems (Zhang & Chen 2022). However, ADVI becomes expensive to compute in high dimensional space when using a full-rank covariance matrix. By using a mean-field approximation, the method can be more efficient, but ignores correlation between parameters and produces biased results (Zhang et al. 2023). To reduce this issue, Zhao & Curtis (2024b) introduced a method called Physically Structured Variational Inference (PSVI), by incorporating spatially localized correlation structures into the variational distribution, which takes into account only correlation of spatial proximity parameters. Nevertheless, all of these ADVI methods can lead to biased posterior estimates due to their implicit Gaussian approximation, for example, for multi-modal posterior distributions (Zhang *et al.* 2021).

Stein variational gradient descent (SVGD) provides a more accurate alternative for probabilistic inference by using a set of tunable particles (representative models) to represent the posterior pdf. By minimizing the KL divergence between the particle distribution and the target distribution, SVGD iteratively reshapes the particles ensemble to align with the posterior probability distribution (Liu & Wang 2016; Leviyev *et al.* 2022). Zhang & Curtis (2020a) applied SVGD to 2-D travel time tomography, and demonstrated that the method can provide accurate and efficient estimates of posterior distributions compared to McMC methods. The method has later been applied to a variety of geophysical problems, including seismic travel time tomography (Zhang & Curtis 2023; Agata *et al.* 2023), earthquake location (Smith *et al.* 2022), hydrogeological inversion (Ramgraber *et al.* 2021), and full waveform inversion (Zhang & Curtis 2020b, 2021; Zhang *et al.* 2023).

Despite the growing interest in variational inference, the method is mainly used to solve 1-D or 2-D problems, and its application to 3-D tomographic problems remains limited (Zhang *et al.* 2021; Zhang *et al.* 2023; Yang *et al.* 2025; Zhao & Curtis 2025). In this study, we use variational inference methods to solve the one-step 3-D surface wave tomographic problems. In the following, we first describe the basic concept of 3-D surface wave tomography and variational inference methods, including ADVI, SVGD and stochastic SVGD (sSVGD, Gallego & Insua 2018). In section 3, we apply the suite of methods to a synthetic 3-D surface wave tomography problem and compare their results to assess performance of those methods. In section 4,we use these methods to study the shear wave velocity structure of the crust and upper mantle in southwest China using Rayleigh wave dispersion data extracted from ambient noise (Liu *et al.* 2021). The results show that variational inference can obtain higher-resolution shear velocity models than those obtained using traditional methods, along with reliable uncertainty estimates. We therefore conclude that variational inference is practical and feasible in solving 3-D high dimensional surface wave tomographic problems, and can be applied fruitfully to image the Earth's subsurface.

## 2. Methods

### 2.1. Variational inference

Bayesian inference calculates the posterior probability density $p(\boldsymbol{m}|\boldsymbol{d}_{obs})$ for model parameters $\boldsymbol{m}$ by revising prior beliefs $p(\boldsymbol{m})$ in light of new data. This revision is governed by Bayes' theorem:

$$p(\boldsymbol{m}|\boldsymbol{d}_{obs}) = \frac{p(\boldsymbol{d}_{obs}|\boldsymbol{m})p(\boldsymbol{m})}{p(\boldsymbol{d}_{obs})} \tag{1}$$

where $p(\boldsymbol{d}_{obs}|\boldsymbol{m})$ is the likelihood which describes the probability of observing data $\boldsymbol{d}_{obs}$ given $\boldsymbol{m}$, $p(\boldsymbol{m})$ represents the prior pdf which describes information that is known independently of data, and $p(\boldsymbol{d}_{obs})$ is a normalization factor called the evidence. The likelihood function is often assumed to be a Gaussian distribution around the data predicted from model $\boldsymbol{m}$ using the known physical relationship, as this is often a reasonable approximation to the pdf of noise or errors in the measured data.

Variational inference tackles Bayesian inference using optimization. The method approximates the posterior pdf $p(\boldsymbol{m}|\boldsymbol{d}_{obs})$ by finding a distribution $q^*(\boldsymbol{m})$ within a predefined family of known probability distributions $Q = \{q(\boldsymbol{m})\}$ that minimizes the KL divergence between $q(\boldsymbol{m})$ and the true posterior pdf:

$$q^*(\boldsymbol{m}) = arg\ min_{q\epsilon Q} KL[q(\boldsymbol{m})||p(\boldsymbol{m}|\boldsymbol{d}_{obs})] \tag{2}$$

The KL divergence measures the difference between two probability distributions and can be expressed as:

$$KL[q(\boldsymbol{m})||p(\boldsymbol{m}|\boldsymbol{d}_{obs})] = \mathbb{E}_q[\log q(\boldsymbol{m})] - \mathbb{E}_q[\log p(\boldsymbol{m}|\boldsymbol{d}_{obs})] \tag{3}$$

where the expectation is taken with respect to the distribution $q(\boldsymbol{m})$. The KL divergence is nonnegative and only equals zero when $q(\boldsymbol{m}) = p(\boldsymbol{m}|\boldsymbol{d}_{obs})$ (Kullback & Leibler 1951). Expanding the posterior pdf using equation (1), the KL divergence becomes:

$$KL[q(\boldsymbol{m})||p(\boldsymbol{m}|\boldsymbol{d}_{obs})] = \mathbb{E}_q[\log q(\boldsymbol{m})] - \mathbb{E}_q[\log p(\boldsymbol{m}, \boldsymbol{d}_{obs})] + \log p(\boldsymbol{d}_{obs}) \tag{4}$$

However, the evidence term $\log p(\boldsymbol{d}_{obs})$ is intractable, as it involves a high-dimensional integral whose evaluation scales exponentially with the number of model parameters. We thus rearrange equation (4) to obtain the evidence lower bound (ELBO):

$$\text{ELBO}[q] = \log p(\boldsymbol{d}_{obs}) - KL[q(\boldsymbol{m})||p(\boldsymbol{m}|\boldsymbol{d}_{obs})] = \mathbb{E}_q[\log p(\boldsymbol{m}, \boldsymbol{d}_{obs})] - \mathbb{E}_q[\log q(\boldsymbol{m})] \tag{5}$$

Since the KL[$q \parallel p$]$\geqslant$0, equation (5) defines a lower bound for the evidence $\log p(\boldsymbol{d}_{obs})$. Because the evidence term is a constant for a given problem, minimizing the KL-divergence is equivalent to maximizing the ELBO. Consequently, variational inference in equation (2) can be expressed as:

$$q^*(\mathbf{m}) = \arg\max_{q\in Q} \text{ELBO}[q(\mathbf{m})] \tag{6}$$

The selection of the variational family $Q$ is critical as it determines both the approximation accuracy and the computational tractability of the optimization problem. Different choices of $Q$ lead to different algorithms. In this study we describe two types of approaches: automatic differentiation variational inference (ADVI), and Stein variational gradient descent (SVGD), including its stochastic variant (sSVGD), and evaluate their performance in 3D surface wave inversion problems.

**2.2 Automatic differential variational inference (ADVI)**

ADVI utilizes Gaussian distributions as variational family (Kucukelbir *et al.* 2017). Since geophysical parameters often exhibit hard constraints (e.g., seismic velocity is always positive), we first use an invertible transform $T$: $\theta = T(\boldsymbol{m})$ to map constrained parameters $\boldsymbol{m}$ into an unconstrained space $\theta$. A commonly-used transform is logit function (Team *et al.* 2016; Zhang & Curtis 2020a):

$$m_i = f(\theta_i) = a_i + \frac{b_i - a_i}{1+\exp(-\theta_i)}$$

$$\theta_i = f^{-1}(m_i) = \log(m_i - a_i) - \log(b_i - m_i) \tag{7}$$

where $m_i$ is the $i^{th}$ model parameter bounded by the lower and upper bounds $a_i$ and $b_i$ respetively, $\theta_i$ is the transformed variable in an unconstrained space. In this unconstrained space the joint probability $p(\boldsymbol{m}, \boldsymbol{d}_{obs})$becomes:

$$p(\theta, \mathbf{d}_{\text{obs}}) = p(\mathbf{m}, \mathbf{d}_{\text{obs}}) \left| \det J_{T^{-1}}(\theta) \right| \tag{8}$$

where $J_{T^{-1}}(\theta)$ is the Jacobian matrix of the inverse of $T$ and | • | denotes absolute value. Define a Gaussian variational family:

$$q(\theta; \boldsymbol{\xi}) = \mathcal{N}(\theta \mid \boldsymbol{\mu}, \boldsymbol{\Sigma}) = \mathcal{N}(\theta \mid \boldsymbol{\mu}, \mathbf{L}\mathbf{L}^{\top}) \tag{9}$$

where $\boldsymbol{\xi}$ represents variational parameters, that is the mean vector $\boldsymbol{\mu}$ and the covariance matrix $\boldsymbol{\Sigma}$. The covariance matrix is reparameterized using a Cholesky factorization $\boldsymbol{\Sigma} = \mathbf{L}\mathbf{L}^{\top}$ with a lower-triangular matrix $\mathbf{L}$ to ensure positive semidefiniteness. The off-diagonal values of $\mathbf{L}$ describe correlation between model parameters. However, a full-rank covariance matrix becomes computationally intractable for high dimensional problems (as in 3D FWI; Zhang *et al.* 2023) as it requires $n(n + 1)/2$ parameters to construct $\mathbf{L}$ where n is the number of parameters. To reduce this issue, a mean-field approximation may be employed by using a diagonal $\mathbf{L}$ (Kucukelbir *et al.* 2017), but this choice tends to strongly underestimate posterior parameter uncertainties because correlation between parameters is neglected (Kucukelbir *et al.* 2017; Zhang *et al.* 2023).

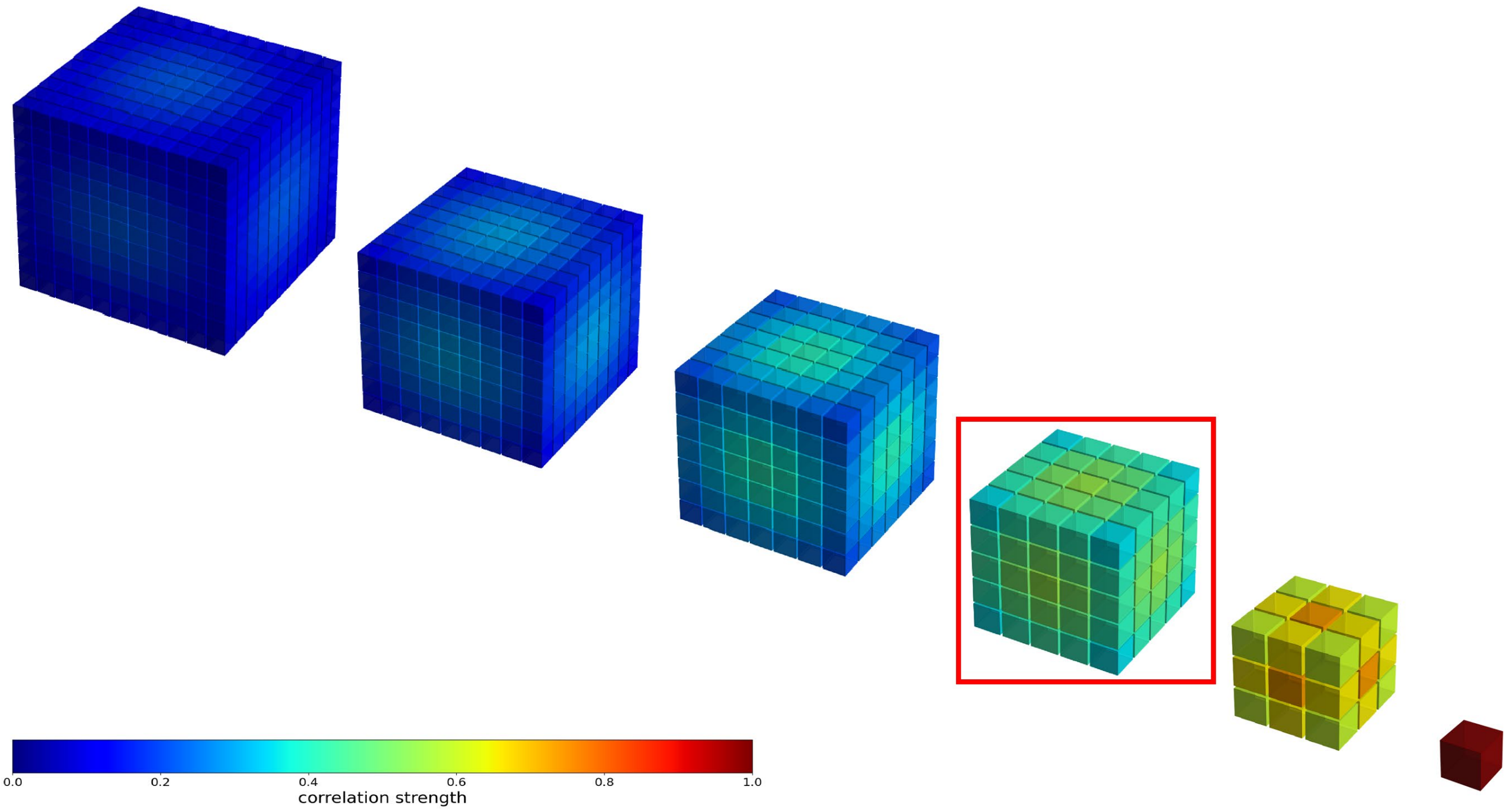

**Figure 1.** The parameter correlation strength in a 3D model discretized into square grid cells (nx×ny×nz), with the central cell (highlighted in red) serving as a reference. The correlation magnitude between the central cell and surrounding cells exhibits an inverse relationship with spatial distance – cells proximal to the central reference show stronger correlation, while those at greater distances display progressively weaker association. This indicates that distal cells maintain only negligible correlation with the central cell.

Alternatively, one can construct an $\mathbf{L}$ which only captures correlation between spatially proximate volumes (Zhao & Curtis 2024b), as in reality the magnitude of correlation between two cells often decreases with their distance apart (Fig. 1). Thus, for an $n$-dimensional problem displayed in Fig. 1 ($n = nx \times ny \times nz$), $\mathbf{L}$ can be set as a sparse matrix to capture only correlation between proximate cells (see Appendix A for details on constructing $\mathbf{L}$). By doing this, the covariance matrix $\mathbf{\Sigma} = \boldsymbol{L}\boldsymbol{L}^T$ only contains correlation between pairs of adjacent cells in each direction (cells surrounding the central red cell in Fig. 1). The method therefore captures the structural correlation of actual physics, which is referred as physically structured variational inference (PSVI), and has been shown to produce more accurate results than mean-field ADVI, and are more computational efficient than full-rank ADVI (Zhao & Curtis 2024b).

With the above definition the variational problem in equation (6) can be written as:

$$\boldsymbol{\xi}^* = \arg\max_{\boldsymbol{\xi}} \mathrm{ELBO}[q(\boldsymbol{\theta};\boldsymbol{\xi})] = \arg\max_{\boldsymbol{\xi}} \mathbb{E}_q[\log p(T^{-1}(\boldsymbol{\theta}), \mathbf{d}_{\mathrm{obs}}) + \log|\det \mathbf{J}_{T^{-1}}(\boldsymbol{\theta})|] - \mathbb{E}_q[\log q(\boldsymbol{\theta};\boldsymbol{\xi})] \quad (10)$$

This optimization problem can be solved by using gradient ascent methods. As shown in Kucukelbir *et al.* (2017) the gradients of the ELBO with respect to $\boldsymbol{\mu}$ and $\mathbf{L}$ are:

$$\nabla_{\boldsymbol{\mu}} \mathrm{ELBO} = \mathbb{E}_{N(\boldsymbol{\eta}|0,\mathbf{I})}\left[\nabla_{\mathbf{m}} \log p(\mathbf{m}, \mathbf{d}_{\mathrm{obs}}) \nabla_{\boldsymbol{\theta}} T^{-1}(\boldsymbol{\theta}) + \nabla_{\boldsymbol{\theta}} \log\left|\det \mathbf{J}_{T^{-1}}(\boldsymbol{\theta})\right|\right] \quad (11)$$

$$\nabla_{\mathbf{L}} \mathrm{ELBO} = \mathbb{E}_{N(\boldsymbol{\eta}|0,\mathbf{I})}\left[\left(\nabla_{\mathbf{m}} \log p(\mathbf{m}, \mathbf{d}_{\mathrm{obs}}) \nabla_{\boldsymbol{\theta}} T^{-1}(\boldsymbol{\theta}) + \nabla_{\boldsymbol{\theta}} \log\left|\det \mathbf{J}_{T^{-1}}(\boldsymbol{\theta})\right|\right)\boldsymbol{\eta}^{\mathrm{T}}\right] + (\mathbf{L}^{-1})^{\mathrm{T}} \quad (12)$$

where $\boldsymbol{\eta}$ is a random variable generated from a standard Normal distribution $N(\mathbf{0}, \mathbf{I})$. The expectations in the above equations can be estimated by Monte Carlo (MC) integration, which in practice only requires a low number of samples because the optimization is performed over many iterations so that statistically the gradients will lead to convergence toward the correct solution (Kucukelbir *et al.* 2017). The final approximation $q^*(\boldsymbol{m})$ is obtained by transforming $q^*(\boldsymbol{\theta})$ back to the original space.

**2.3 Stein variational gradient descent (SVGD)**

SVGD approximates the target posterior pdf using a set of models (called particles) $\{\boldsymbol{m}_i\}_{i=1}^n$. In SVGD, these particles are transported toward the posterior distribution by iteratively minimizing the KL divergence between the particle distribution and the posterior pdf (Liu & Wang 2016). Since the distribution of a set of particles is in principle entirely flexible, this method can provide more accurate results than ADVI (Zhang & Curtis 2020a). SVGD updates particles using a smooth perturbation:

$$T(\boldsymbol{m}_i) = \boldsymbol{m}_i + \epsilon\boldsymbol{\phi}(\boldsymbol{m}_i) \tag{13}$$

where $\boldsymbol{\phi} = [\phi_1, \ldots, \phi_{\mathrm{n}}]$ is a smooth vector function that describes the perturbation direction and $\epsilon$ is the magnitude of the perturbation. Assume $T$ is invertible and define $q_T(\boldsymbol{m})$ as the transformed probability distribution of pdf $q(\mathbf{m})$. The gradient of the KL-divergence between $q_T$ and the posterior pdf $p$ with respect to $\epsilon$ can be computed as (Liu & Wang 2016):

$$\nabla_\epsilon \mathrm{KL}[q_T \,\|\, p]\big|_{\epsilon=0} = -\mathrm{E}_q\left[\mathrm{trace}\left(A_p\phi(\mathbf{m})\right)\right] \tag{14}$$

where $A_p$ is the Stein operator defined by $A_p\phi(\mathbf{m}) = \nabla_{\mathbf{m}} \log p(\mathbf{m})\phi(\mathbf{m})^T + \nabla_{\mathbf{m}}\phi(\mathbf{m})$ . According to equation (14) we can obtain the steepest descent direction of the KL-divergence by maximizing the right-hand expectation, and consequently the KL divergence can be minimized by iteratively stepping a small distance in that direction.

Liu & Wang (2016) showed that the optimal $\boldsymbol{\phi}^*$ which maximizes the expectation in equation (14) can be derived using kernel functions. Say $x, y \in X$ and define a mapping $\psi$ from $X$ to a space where an inner product $\langle , \rangle$ is defined (called a Hilbert space); *a kernel* is a function that satisfies $k(x, y) = \langle \psi(x), \psi(y) \rangle$ . Assume a kernel $k(\mathbf{m}', \mathbf{m})$, the optimal $\boldsymbol{\phi}^*$ can be expressed as (Liu & Wang 2016):

$$\Phi^* \propto E_{\{\mathbf{m}' \sim q\}}\left[A_p k(\mathbf{m}', \mathbf{m})\right] \tag{15}$$

where the expectation can be approximated using the particles mean. The KL divergence can therefore be minimized by iteratively applying the transform $T(\boldsymbol{m}_i) = \boldsymbol{m}_i + \epsilon\boldsymbol{\phi}^*(\boldsymbol{m}_i)$ to a set of initial particles $\{\boldsymbol{m}_i^0\}$:

$$\phi_l^*(\mathbf{m}) = \frac{1}{n}\sum_{j=1}^{n}\left[k(\mathbf{m}_j^l, \mathbf{m})\nabla_{\mathbf{m}_j^l} \log p(\mathbf{m}_j^l \,|\, \mathbf{d}_{\mathrm{obs}}) + \nabla_{\mathbf{m}_j^l} k(\mathbf{m}_j^l, \mathbf{m})\right] \tag{16}$$

$$\mathbf{m}_i^{l+1} = \mathbf{m}_i^l + \epsilon^l \phi_l^*(\mathbf{m}_i^l) \tag{17}$$

where superscript $l$ denotes the $l^{th}$ iteration and $n$ is the number of particles. If the step size $\{\epsilon^l\}$ is sufficiently small then the transform $T$ is invertible, and the process converges to the posterior pdf asymptotically as the number of particles tends to infinity.

For the kernel function we use a commonly-used radial basis function (RBF):

$$k(\mathbf{m}, \mathbf{m}') = \exp\left[-\frac{\|\mathbf{m} - \mathbf{m}'\|^2}{2h^2}\right] \tag{18}$$

where $h$ is a bandwidth parameter that controls the intensity of particle interactions depending on distance. As suggested by several studies (Liu & Wang 2016; Zhang & Curtis 2020a,b), we choose $h = \tilde{d}/\sqrt{2\log n}$ where $\tilde{d}$ is the median of pairwise distances between all particles. This choice ensures balanced contributions between each particle's local gradient and gradients from all other particles. Note that for the RBF kernel, the second term of $\boldsymbol{\phi}^*$ in equation (16) becomes $\sum_j \frac{\mathbf{m}-\mathbf{m}_j}{\mathrm{h}^2} k(\mathbf{m}_j, \mathbf{m})$ which drives the particle $\mathbf{m}$ away from its neighbouring particles when the kernel takes high values. This second term therefore acts as a repulsive force which prevents the particles from collapsing to a single mode, whereas the first term consists of kernel weighted gradients which drives the particles toward high probability areas.

**2.4 Stochastic SVGD**

Despite its wide applications (Gong *et al.* 2019; Zhang & Curtis 2020a,b; Pinder *et al.* 2020), SVGD can produce biased estimations in high-dimensional space due to finite-particle implementation and computational constraints (Ba *et al.* 2021). To improve accuracy, Gallego & Insua (2018) proposed a variant of SVGD, called stochastic SVGD (sSVGD), which transforms SVGD into an McMC method with multiple interacting chains by introducing a Gaussian noise term to the dynamics of SVGD. By doing this one can start collecting many samples that represent the posterior pdf after a burn-in period instead of having to use a large number of particles from the beginning. In addition, this method guarantees asymptotic convergence as the number of iterations tends to infinity, which cannot be achieved by SVGD with finite number of particles (Leviyev *et al.* 2022).

To introduce the sSVGD algorithm, we start from a stochastic differential equation (SDE):

$$d\mathbf{z} = \mathbf{f}(\mathbf{z})dt + \sqrt{2\mathbf{D}(\mathbf{z})}\,d\mathbf{W}(t) \tag{19}$$

where $\mathbf{f}(\mathbf{z})$ is called the *drift*, $\mathbf{W}(t)$ is a Wiener process, and $\mathbf{D}(\mathbf{z})$ is a positive semidefinite diffusion matrix. Generally, all continuous Markov processes can be expressed as a SDE of the above form (Oksendal 2013). If we denote the posterior distribution as $p(\mathbf{z})$, Ma *et al.* (2015) proposed a SDE that converges to the distribution $p(\mathbf{z})$:

$$f(\mathbf{z}) = \left[\mathbf{D}(\mathbf{z}) + \mathbf{Q}(\mathbf{z})\right]\nabla \log p(\mathbf{z}) + \Gamma(\mathbf{z}) \tag{20}$$

where $\boldsymbol{Q}(\mathbf{z})$ is a skew-symmetric curl matrix, and $\Gamma_i(\mathbf{z}) = \sum_{j=1}^{d} \frac{\partial}{\partial \mathbf{z}_j}\left(\mathbf{D}_{ij}(\mathbf{z}) + \mathbf{Q}_{ij}(\mathbf{z})\right)$ is a correction term.

By discretizing the SDE using Euler–Maruyama, we obtain a practical algorithm:

$$\mathbf{z}_{t+1} = \mathbf{z}_t + \epsilon_t \left[ (\mathbf{D}(\mathbf{z}_t) + \mathbf{Q}(\mathbf{z}_t)) \nabla \log p(\mathbf{z}_t) + \Gamma(\mathbf{z}_t) \right] + \mathcal{N}(\mathbf{0}, 2\epsilon_t \mathbf{D}(\mathbf{z}_t)) \tag{21}$$

where $\mathcal{N}(\mathbf{0}, 2\epsilon_t \mathbf{D}(\mathbf{z}_t))$ represents a Gaussian distribution. The gradient $\nabla \log p(\mathbf{z}_t)$ can be computed using the entire data set, or uniformly randomly selected mini-batches of the data which produces a stochastic gradient approximation. In either case the above process converges to the posterior distribution asymptotically as $\epsilon_t \to 0$ and $t \to \infty$ (Ma *et al.* 2015). Matrix $\boldsymbol{D}(\mathbf{z})$ and $\boldsymbol{Q}(\mathbf{z})$ can be adjusted to obtain faster convergence to the posterior distribution. For example, by setting $\boldsymbol{D} = \mathbf{I}$ and $\boldsymbol{Q} = \mathbf{0}$ one obtains the stochastic gradient Langevin dynamics algorithm (Welling & Teh 2011). If we augment the state space $\mathbf{z}$ with a moment term $\mathbf{x}$ to obtain an augmented space $\bar{\boldsymbol{z}} = (\boldsymbol{z}, \boldsymbol{x})$, and set $\mathbf{D} = \mathbf{0}$ and $\boldsymbol{Q} = \begin{bmatrix} \mathbf{0} & -\mathbf{I} \\ \mathbf{I} & \mathbf{0} \end{bmatrix}$, the stochastic Hamiltonian Monte Carlo (HMC) method can be derived (Chen *et al.* 2014).

In sSVGD, for the set of particles $\{\boldsymbol{m}_i\}$ defined in the above section, we construct an augmented space $\mathbf{z} = (\boldsymbol{m}_1, \boldsymbol{m}_2, \ldots, \boldsymbol{m}_n) \in \mathbb{R}^{nd}$ that concatenates the $n$ particles, and use equation (21) to obtain a valid sampler that runs multiple ($n$) interacted chains in parallel to generate samples from the posterior distribution $p(\boldsymbol{z}) = \prod_{i=1}^{n} p(\boldsymbol{m}_i | \boldsymbol{d}_{obs})$. Define a matrix $\boldsymbol{K}$:

$$K = \frac{1}{n} \begin{bmatrix} k(\mathbf{m}_1, \mathbf{m}_1)\mathbf{I}_{d\times d} & \cdots & k(\mathbf{m}_1, \mathbf{m}_n)\mathbf{I}_{d\times d} \\ \vdots & \ddots & \vdots \\ k(\mathbf{m}_n, \mathbf{m}_1)\mathbf{I}_{d\times d} & \cdots & k(\mathbf{m}_n, \mathbf{m}_n)\mathbf{I}_{d\times d} \end{bmatrix} \tag{23}$$

where $k(\boldsymbol{m}_i, \boldsymbol{m}_j)$ is a kernel function and $I_{d\times d}$ is an identity matrix. According to the definition of kernel functions, the matrix $\boldsymbol{K}$ is positive definite (Gallego & Insua 2018). Let $\boldsymbol{D_k} = \boldsymbol{K}, \boldsymbol{Q_k} = \mathbf{0}$, equation (21) becomes:

$$z_{t+1} = z_t + \epsilon_t \left[ \mathbf{K} \nabla \log p(\mathbf{z}_t) + \nabla \cdot \mathbf{K} \right] + \mathcal{N}(0, 2\epsilon_t \mathbf{K}) \tag{24}$$

Comparing this equation with equations (16) and (17), SVGD can be regarded as a special case of equation (24) without Gaussian noise term. This equation therefore defines a stochastic gradient McMC method with SVGD gradients, which is called stochastic SVGD (sSVGD). According to the discussion above, this process converges to the posterior distribution $p(\mathbf{z}) = \prod_{i=1}^{n} p(\boldsymbol{m}_i | \boldsymbol{d}_{obs})$ asymptotically. Note that when the number of particles is large enough, the noise term would be tiny according to equation (23). Consequently in such case the method produces the same results as standard SVGD.

In order to use equation (24) to sample the posterior distribution, we need to draw samples from the Gaussian distribution $\mathcal{N}(0, 2\epsilon_t \mathbf{K})$. This requires computing the lower triangular Cholesky decomposition of the $nd \times nd$ matrix $\boldsymbol{K}$, which can be computationally expensive. To compute the noise term efficiently, we define a block-diagonal matrix $\boldsymbol{D_k}$:

$$\mathbf{D_K} = \frac{1}{n}\begin{bmatrix} \overline{\mathbf{K}} & & \\ & \ddots & \\ & & \overline{\mathbf{K}} \end{bmatrix} \tag{25}$$

where $\overline{\mathbf{K}}$ is a $n \times n$ matrix with $\overline{\boldsymbol{K}_{ij}} = k(\boldsymbol{m}_i, \boldsymbol{m}_j)$. Notice that with this definition, $\boldsymbol{D_k}$ can be constructed from $\boldsymbol{K}$ using $\boldsymbol{D_k} = \mathbf{PKP}^{\mathrm{T}}$ where $\mathbf{P}$ is a permutation matrix:

$$\mathbf{P} = \begin{bmatrix} 1 & & & & & & & & & \\ & & & 1 & & & & & & \\ & & & & & & \ddots & & & \\ - & - & - & - & - & - & - & \underline{1} & - & - \\ & 1 & & & & & & & & \\ & & & & 1 & & & & & \\ & & & & & & \ddots & & & \\ - & - & - & - & - & - & - & - & \underline{1} & - \\ & \ddots & & & \ddots & & \ddots & & \ddots & \\ - & - & \overline{1} & - & - & - & - & - & - & - \\ & & & & & 1 & & & & \\ & & & & & & \ddots & & & \\ & & & & & & & & & 1 \end{bmatrix} \tag{26}$$

This permutation matrix transforms vector $\mathbf{z}$ from particle-major ordering (sequential particle listing) to dimension-major ordering (first coordinates of all particles, followed by second coordinates, etc.). The noise term η can therefore be generated using:

$$\begin{aligned} \eta &\sim \mathcal{N}(0, 2\epsilon_t \mathbf{K}) \\ &\sim \sqrt{2\epsilon_t}\,\mathbf{P}^{\mathrm{T}}\mathbf{P}\mathcal{N}(0, \mathbf{K}) \\ &\sim \sqrt{2\epsilon_t}\,\mathbf{P}^{\mathrm{T}}\mathcal{N}(0, \mathbf{D_K}) \\ &\sim \sqrt{2\epsilon_t}\,\mathbf{P}^{\mathrm{T}}\mathbf{L}_{\mathbf{D_K}}\mathcal{N}(0, \mathbf{I}) \end{aligned} \tag{27}$$

Given the block-diagonal structure of $\boldsymbol{D_k}$, its Cholesky factor $\boldsymbol{L_{D_k}}$ can be calculated easily as we only need to calculate the lower triangular Cholesky decomposition of matrix $\overline{\mathbf{K}}$. Since the number of particles $n$ is typically modest in practice, evaluation of the noise term is computationally negligible. Equation (24) can therefore be directly employed to sample the posterior distribution.

**2.5 3D variational surface wave tomography**

In order to use variational inference methods in 3D surface wave tomography, it is required to efficiently compute travel times and their gradients with respect to model parameters at varying periods along arbitrary source-receiver paths. Ideally a fully 3-D wavefield simulation (e.g., spectral-element method) could be used but these are generally computationally too expensive (Chen *et al.* 2014; Gao & Shen 2014). To balance accuracy and computational efficiency, we adopt a ray tracing-based approximation method (Fang *et al.* 2015), which first calculates phase or group velocity dispersion curves at each geographic point using the 1-D velocity profile beneath this point, and then computes source-receiver travel times at each period using the calculated phase or group velocity map at this period. Under this approximation, for a logarithm posterior pdf function $\mathbf{F}$ (e.g., $\mathbf{F} = \log p(\boldsymbol{m}, \boldsymbol{d}_{obs})$ as in equation 12), the gradient $\nabla_{\boldsymbol{m}}\mathbf{F}$ can be calculated using chain rule:

$$\nabla_{\boldsymbol{m}}\mathbf{F} = \sum_{i}^{N}\sum_{\tau}^{T} \frac{\partial F_{i,\tau}}{\partial t_i(\tau)} \frac{\partial t_i(\tau)}{\partial C_i(\tau)} \frac{\partial C_i(\tau)}{\partial \boldsymbol{m}} \tag{28}$$

where N and T represent the number of source-receiver paths and discrete periods in surface wave dispersion measurements, respectively. $\frac{\partial F_{i,\tau}}{\partial t_i(\tau)}$ quantifies the sensitivity of the function $\mathbb{F}$ to the $i_{\text{th}}$ phase travel-time perturbations at each measurement period. $\frac{\partial t_i(\tau)}{\partial C_i(\tau)}$ is calculated using the fast marching method (Rawlinson & Sambridge 2004), and $\frac{\partial C_i(\tau)}{\partial \boldsymbol{m}}$ represents the Frechét kernel of phase velocity with respect to model parameters (e.g., shear wave velocity), computed using matrix propagator methods (Herrmann 2013).

## 3. Synthetic tests

We first apply the suite of methods to a 3D synthetic example. The true model is composed of three homogeneous layers with shear wave velocities of 3.0, 4.0 and 5.0 km/s, respectively. In the second layer we constructed a square high-speed anomaly (4.5 km/s) and a low-speed anomaly (3.5 km/s) in two specified locations (Fig. 2a and b). The P-wave velocity ($v_p$) is determined by a fixed $\frac{v_p}{v_s}$ ratio with a typical crustal value of 1.73. Density $\rho$ is expressed as a function of $v_p$ through an empirical relationship: $\rho = 2.35 + 0.036 \times (v_p - 3.0)^2$ (Kurita 1973). We deployed 173 receivers whose locations are set as the actual station distribution in Southwest China (Liu *et al.* 2021). Each receiver is also treated as a virtual source to simulate a typical ambient noise tomography (Shapiro *et al.* 2005). We discretized the velocity model using a 27×23×71 grid, and calculated the phase velocity travel-times of Rayleigh waves at periods from 5 to 50 s between each station pair using the forward method described above. In addition, two percent Gaussian noise is added to the data. This produces a total of 14,878 inter-receiver travel-times for each period.

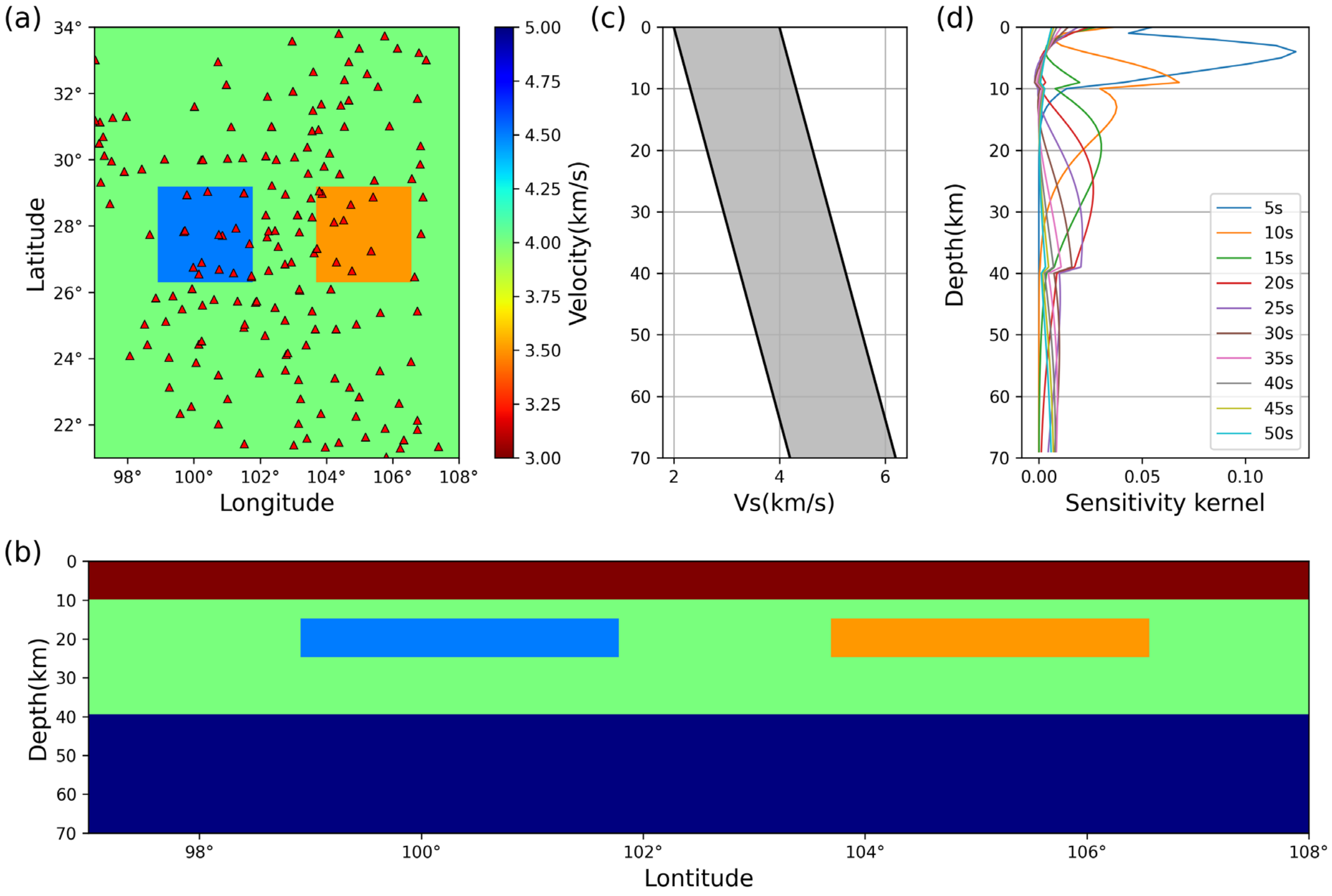

**Figure 2.** True model used for the synthetic test. Red triangles show the locations of receivers. (a) Horizontal section of the true model at depth of 20 km; (b) Vertical section at latitude=27.5°N; (c) Prior distribution used in the inversion: a Uniform distribution with a width of 2.2 km/s at each depth; (d) Phase velocity sensitivity kernels calculated for a 1D profile with the three homogeneous layers (without those anomalies in the second layer).

For variational inversion, the velocity model **m** is parameterized on a regular 0.5°×0.5° horizontal grid, with vertical intervals of 5 km between 0 and 40 km depth and 10 km between 40 and 70 km depth. The noise level is fixed to be 2% of travel-times for all inversions. The prior pdf of the velocity in each cell is set to be a Uniform distribution over an interval of 2.2 km/s at each depth to encompass the true model (Fig. 2c). In addition, we impose spatial smoothness constraints on model parameters. To achieve this, we define a second-order finite-difference operator **S** that calculates the curvature of model parameters between adjacent grid cells in the 3D velocity model. By assuming a Gaussian distribution for the product **Sm**, we enforce smoothness through:

$$p(\mathbf{Sm}) \propto \exp\left(-\frac{1}{2\sigma_s^2}(\mathbf{Sm})^T(\mathbf{Sm})\right) \tag{29}$$

where $\sigma_s$ is a hyperparameter controlling the strength of spatial smoothing. Equation (29) can be interpreted as applying Tikhonov regularization with matrix **S** to **m** (Golub *et al.* 1999). By combining with the Uniform prior distribution $p(\mathbf{m})$, the resulting smoothed prior pdf can be expressed as:

$$P_{\text{smooth}}(\mathbf{m}) \propto p(\mathbf{Sm})p(\mathbf{m}) \tag{30}$$

Here we choose different $\sigma_s$ for different directions, that is, $\sigma_x$=0.2, $\sigma_y$=0.2, $\sigma_z$=1.0 in the latitude, longitude, and depth direction, respectively.

For mean field ADVI and PSVI, we set the initial Gaussian distribution in the unconstrained space to be a standard Normal distribution $N(\theta|0, I)$. In PSVI we employ a 5×5×5 correlation kernel to model the main correlation between model parameters (see Fig. 1 and Appendix A). The initial distribution is then updated using the ADAM algorithm (Kingma & Ba 2014) for 2,000 iterations for each method. At each iteration the gradients are calculated using eight Monte Carlo samples. The obtained probability distribution is transformed back to the original space to obtain the final approximation, from which we generate 1,000 samples to compute the mean and standard deviation to visualize the results.

For sSVGD we start from 40 particles that are generated from the prior distribution, and transform them to the unconstrained space using equation (7). Those particles are then updated (sampled) using equation (24) for 2,000 iterations with a burn-in period of 1,000. To reduce the memory and storage cost, we only retain every second sample after the burn-in period. This results in a total of 20,000 samples, which are transformed back to the original space to calculate statistics of the estimated posterior pdf.

**3.1 Model comparison**

Fig. 3 shows horizontal profiles of the results at the depth of 20 km obtained using the suite of methods. Overall, the mean models capture the true structure, including the high- and low-velocity anomalies in the second layer. However, the velocity magnitudes are slightly different — the velocity of the high-velocity anomaly is lower ($\approx$4.4 km/s) than the true value (4.5 km/s), and the velocity of the low-velocity anomaly is higher ($\approx$3.6 km/s) than the true value (3.5 km/s). This may reflect uncertainty of the problem itself or may be caused by the smooth prior constraint (see discussion below and Appendix B). The standard deviation models obtained using the three methods generally exhibit lower uncertainty in the central region and higher uncertainty along the periphery as expected, which reflects the insufficient data coverage outside the station array. In addition, the results obtained using mean-field ADVI show smaller uncertainty than those from PSVI and sSVGD because of absence of correlation in mean-field ADVI (Blei *et al.* 2017; Zhang *et al.* 2021). Moreover, while mean-field ADVI yields generally low standard deviation within the array, PSVI and sSVGD show more complex features. For example, both results obtained using PSVI and sSVGD exhibits relatively lower standard deviation at the location of the high velocity anomaly because of relatively denser station distribution, or may be caused by the fact that rays concentrate at high velocity anomalies. Note that there are small scale structures in all three standard deviation models which likely reflects the uncertainty of the problem itself.

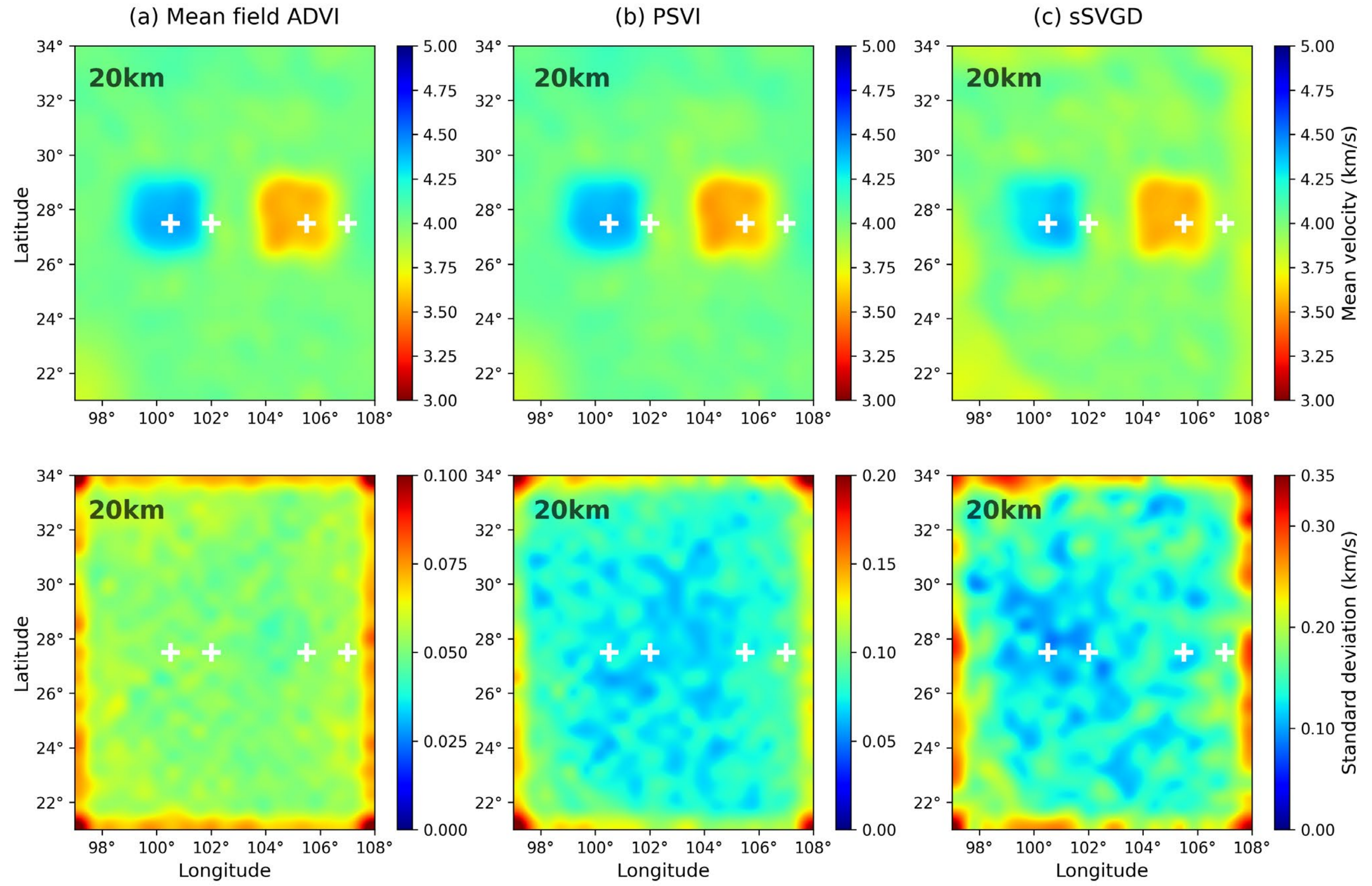

**Figure 3.** The mean (top row) and standard deviation (bottom row) at the horizontal slice of depth = 20 km obtained using (a) mean field ADVI, (b) PSVI, and (c) sSVGD, respectively. The white pluses show the point location which are referred to in the text.

Fig. 4 shows the results at the vertical profile of latitude 27.5°N. Overall, the mean models show similar structure to that of the true model, including the three layers and the two rectangle anomalies within the second layer. The top layer is clearly recovered due to the high sensitivity of surface waves at shallow depth (Fig. 2d), except that the layer boundary is deeper at two sides because of low data coverage in those areas. Within the second layer, the location of the low velocity anomaly is slightly deeper than that of the high velocity anomaly in all three results. Given that the methods are quite different in their theoretical foundation, the results likely reflect the uncertainty of the problem itself. The standard deviation models generally show increasing uncertainty with depth, which reflects the decreasing sensitivity of surface waves when depth increases. Note that the uncertainty is lower in the bottom layer, which is likely due to the cumulative sensitivity at longer periods for depths beyond 70 km because of the imposed half-space boundary at this depth. Similar phenomenon has also been observed in several previous studies (Bodin *et al.* 2012, Zhang *et al.* 2018). There are also high uncertainties around 10km and 40 km depth across all standard deviation models, which is consistent with the rapid decrease in the sensitivity kernels of surface waves across different periods (Fig. 2d).

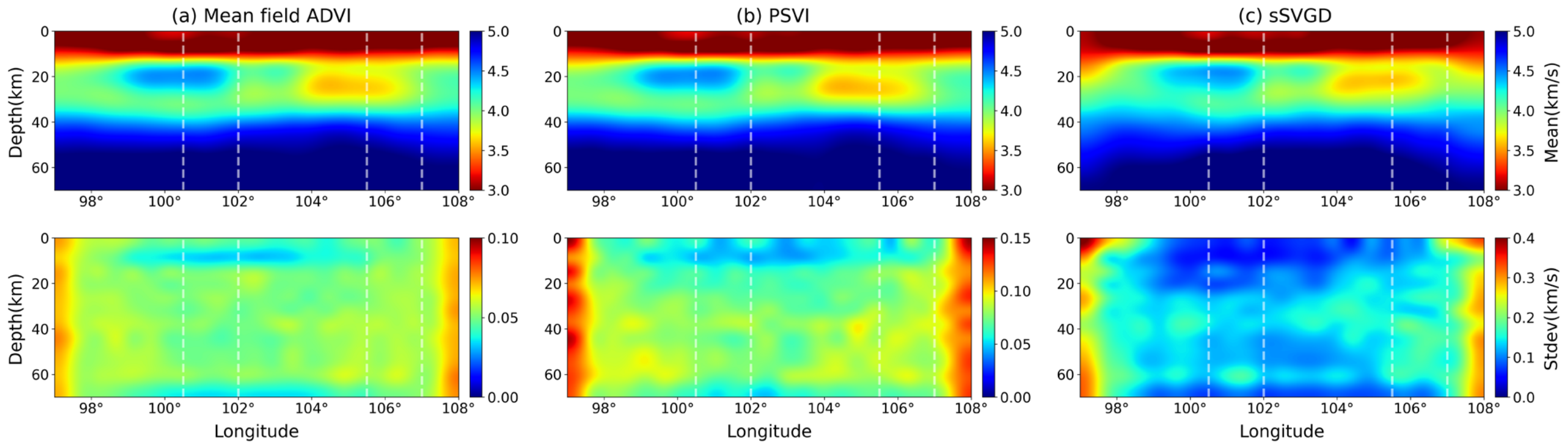

**Figure 4.** The mean (top row) and standard deviation (bottom row) at the vertical slice of Latitude=27.5°N obtained using (a) mean field ADVI, (b) PSVI, and (c) sSVGD, respectively. White dashed lines denote the 1D profile location referred to in the main text.

To further evaluate the results, in Fig. 5 we compare the posterior marginal distributions derived from the three methods along four vertical profiles (white pluses in Fig. 3 and white dashed lines in Fig. 4). Overall, the marginal distributions obtained using the three methods show similar features. For example, all results show generally high probability around the true value. However, there exists details that differ in the results. For example, the distribution obtained using sSVGD are generally wider than those from ADVI and PSVI, which suggests that ADVI and PSVI may underestimate the uncertainty because of their Gaussian assumption (Zhang & Curtis 2020b). In addition, along the third profile between depths of 15–20 km and 25–30 km, the true velocity values fall outside the high-probability regions in the results from ADVI and PSVI (Fig. 5a and 5b), which further indicates that these methods can underestimate uncertainty. By contrast, sSVGD produces more reasonable uncertainty estimates as the distribution generally encompasses the true value within regions of non-zero probability. At around 40 km depth, the true velocity lies outside the high-probability region in all results, which is consistent with the high uncertainties shown in Fig. 4. This is likely caused by the lower sensitivity at this depth and the smoothness prior constraints (see Appendix B).

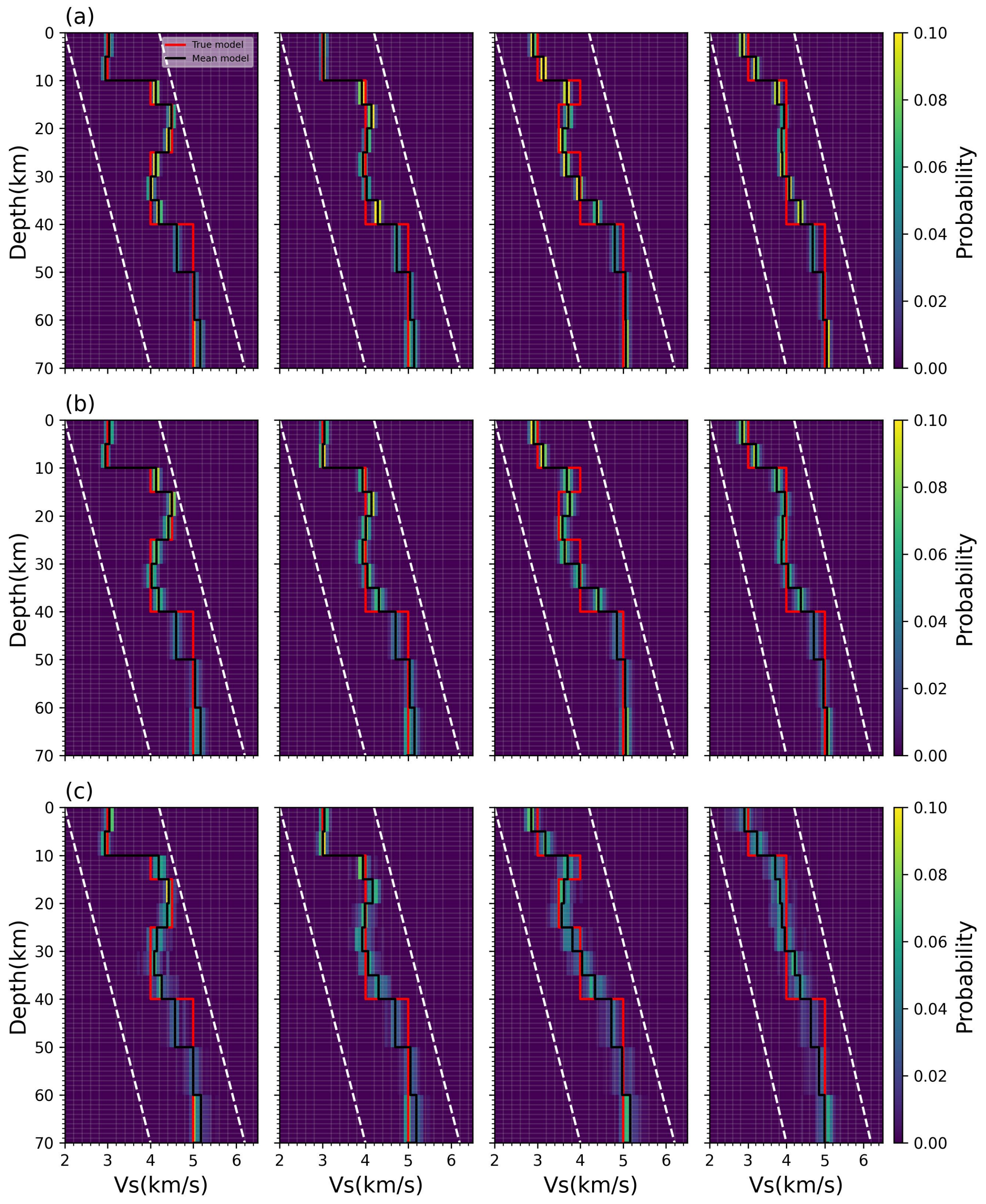

**Figure 5.** The marginal distributions at four vertical profiles obtained using (a) mean field ADVI, (b) PSVI and (c) sSVGD respectively. The locations of these four profiles are represented by white pluses and dashed lines in Figure 3 and 4. White dashed lines show the lower and upper bound of the prior distribution, and black and red lines show the mean and true velocity values.

Strong prior information tends to favor overly smoothed velocity models, as strong smoothness constraints limit the velocity differences between adjacent grid cells. To study effects of the smoothness prior constraints, we compare results using both smoothed and non-smoothed prior pdfs for each method, and analyze their respective consequences.

The results show that the mean velocity models obtained using smoothed prior pdfs exhibit greater spatial continuity with reduced small-scale variations compared to those obtained using non-smoothed prior pdfs (Fig. B1 in appendix B), and the standard deviations obtained using smoothed prior pdfs show lower uncertainty as large velocity contrasts between neighbouring cells are precluded by the prior information. In addition, the results obtained using non-smoothed prior pdfs show a clear low-uncertainty anomaly at the location of the high-velocity anomaly, which becomes less pronounced after applying smoothing constraints. This is likely because the Gaussian smoothing enhances spatial correlations between adjacent cells, and thereby reduces contrast in standard deviation values across velocity boundaries. Similarly, the posterior marginal pdfs obtained using smoothed prior pdfs display narrower probability ranges (Fig. B2).

However, from a Bayesian perspective, without prior knowledge of the true Earth's smoothness, there is no objective reason to prefer any one of these posterior distributions – each reflects a different aspect of posterior uncertainty. While prior intuition may suggest that the models in the top panels of Fig. B1 are under-smoothed, establishing an appropriate prior probability distribution over smoothness levels remains challenging. In such cases, one may try to perform a set of inversions with different smoothness levels, each of which represents a prior belief of smoothness of the subsurface.

### 3.2 Computational cost

Table 1 summarizes the number of simulations and the CPU hours required for each method. The number of simulations serves as a reliable metric for comparing overall computational cost, since forward simulations constitute the most time-consuming component in all methods. Given that all approaches can be fully parallelized, we also report wall time to provide further insight into their practical computational requirements.

**Table 1.** A comparison of computational cost for the 3 inference methods.

| Method | Number of simulations | CPU hours[a] | Wall time (hours) |
|---|---|---|---|
| Mean field ADVI | 16000 | 477.74 | 23.89 |
| PSVI | 16000 | 586.09 | 29.30 |
| sSVGD | 80000 | 1633.79 | 65.35 |

a The CPU used in this study is Intel Xeon Gold 6258R.

The results indicate that mean-field ADVI is the least computationally expensive method, which required 16,000 simulations with 477.74 CPU hours; however, as demonstrated above, it tends to produce systematically biased results. PSVI is nearly as efficient as mean-field ADVI, which took the same number of simulations with slightly more CPU hours (586.09) because of larger covariance matrix computation, yet it improves accuracy and delivers uncertainty estimates comparable to those of sSVGD. This suggests that PSVI method can be useful for obtaining initial, relatively rapid estimates of subsurface structure. sSVGD took the most computational cost, but likely produces the most accurate results as it does not make Gaussian assumption as in mean-field ADVI and PSVI. Note that the above comparison depends on subjective assessments of the point of convergence for each method, so the absolute computational time may not be entirely accurate. Nonetheless, the results offer a reasonable indication of the relative efficiency of each method.

## 4. Application to Southwest China

In this section, we apply the above methods to data collected in Southwest China at the southeastern margin of the Tibetan plateau, which plays an important role for transferring lithospheric materials extruded from the Tibetan plateau to Southeast Asia. This region is characterized by several major fault systems, including the Longmenshan thrust fault (LMSF), the Xianshuihe (XSHF) – Anninghe – Zemuhe – Xiaojiang (XJF) sinistral strike-slip fault system, and the Red River dextral strike-slip fault (RRF) (Fig. 6). These major fault systems accommodate most of the large earthquakes in the region and separate the area into different tectonic units, such as the lens-shaped Chuandian block (bounded by XSHF, XJF, and RRF) (Kan *et al.* 1977; Wang *et al.* 1998), the Songpan-Ganzi block (SGB), the Yangtze craton (YZC), the Qiangtang block (QB), the Lhasa block (LB) and the Shan Thai block (STB) (Fig. 6).

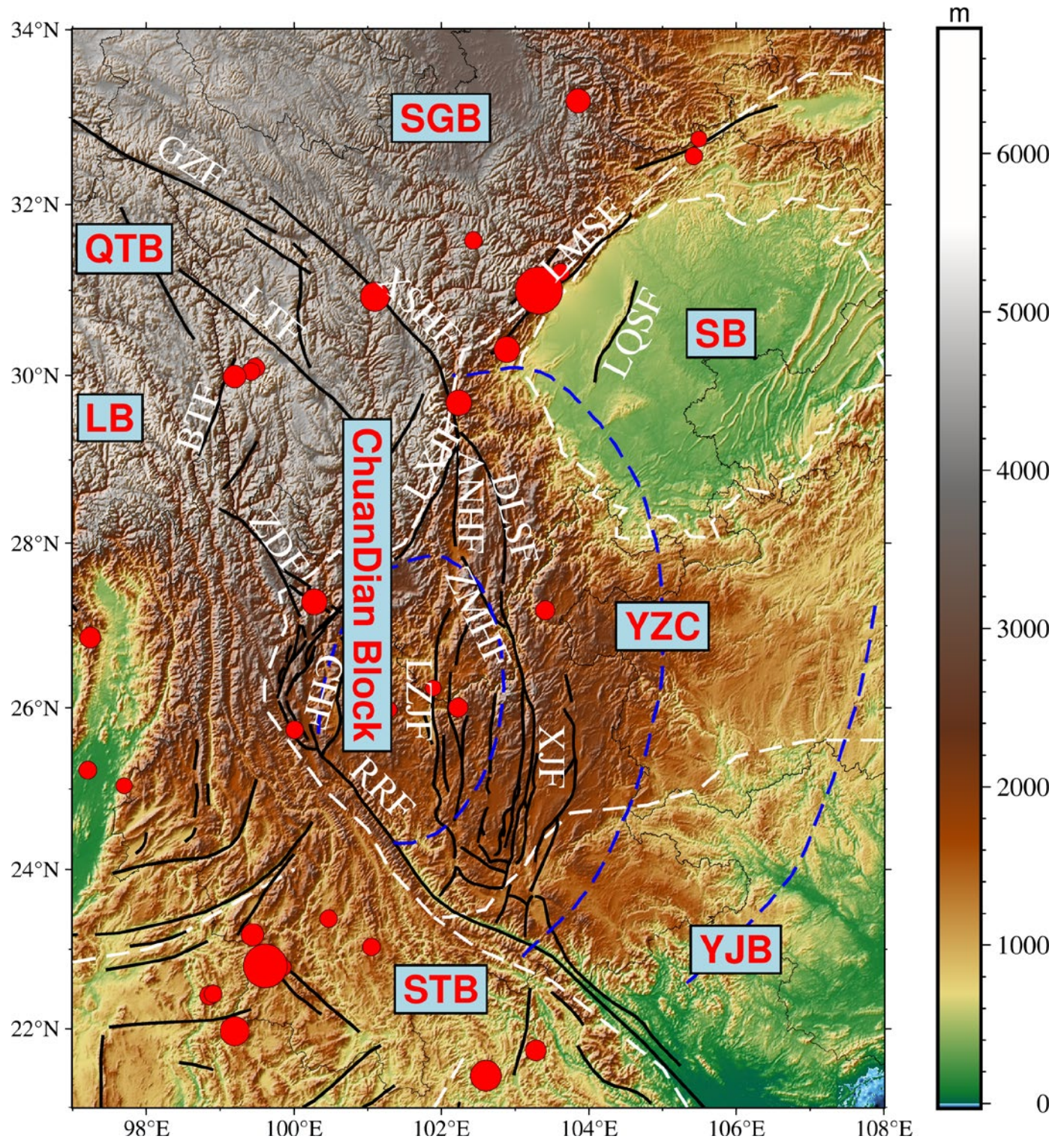

**Figure 6.** Tectonic elements and fault systems in the study region. The dashed white lines indicate tectonic block boundaries. LB, Lhasa block; QTB, Qiangtang block; SB, Sichuan basin; SGB, Songpan-Ganzi block; STB, Shan Thai block; YJB, Youjiang block; YZC, Yangtze craton. The thick solid black lines show major faults. ANHF, Anninghe fault; BTF, Batang fault; GZF, Ganzi fault; LTF, Litang fault; LJXJF, Lijiang-Xiaojinhe fault; LZJF, Lüzhijiang fault; RRF, Red River fault; XJF, Xiaojiang fault; XSHF, Xianshuihe fault; ZDF, Zhongdian fault; ZMHF, Zemuhe fault; DLSF, Daliangshan fault; CHF, Chenghai fault; LQSF, Longquanshan fault. The red filled circles show epicenters of earthquakes (Mw > = 6.0) between 1980 and 2025. The blue dashed lines outline the boundary between the inner, intermediate, and outer zones of the Emeishan large igneous province (ELIP) (Xu *et al.* 2004).

A number of three-dimensional seismic velocity models have been proposed for southwestern China using various types of geophysical data. These include body wave tomography (Huang *et al.* 2015; Xin *et al.* 2019), surface wave tomography (Yao *et al.* 2010; Shen *et al.* 2016; Qiao *et al.* 2018), and joint inversion of surface wave dispersion and receiver functions (Liu *et al.* 2014; Bao *et al.* 2015). However, these studies did not provide reliable uncertainty estimates for velocity models, which introduces difficulties for interpretation. In this study, we apply the above variational inference surface wave tomography to construct shear wave velocity models for this region, together with accurate uncertainty estimates, so that reliability of the results can be better assessed.

### 4.1 Data and inversion details

We use 8100 Rayleigh-wave phase-velocity dispersion curves over periods ranging from 5 to 50 seconds extracted from ambient noise interferometry (Fig. 7b). The dataset integrates measurements from three independent studies (Liu *et al*. 2021): 6910 Rayleigh wave phase-velocity dispersion curves (5–50 s) derived from ambient noise analysis at 124 permanent stations of the China National Seismic Network (Yang *et al.* 2020; Zheng *et al.* 2010); 544 curves (5–45 s) from 35 temporary stations deployed across southwest China by MIT and Lehigh University (Yao *et al*. 2010); and 640 curves (5–40 s) from 16 temporary stations in northern Vietnam and 49 permanent stations in Yunnan Province, installed by the Institute of Earth Sciences, Academia Sinica (Qiao *et al.* 2018; Nguyen *et al.* 2013).

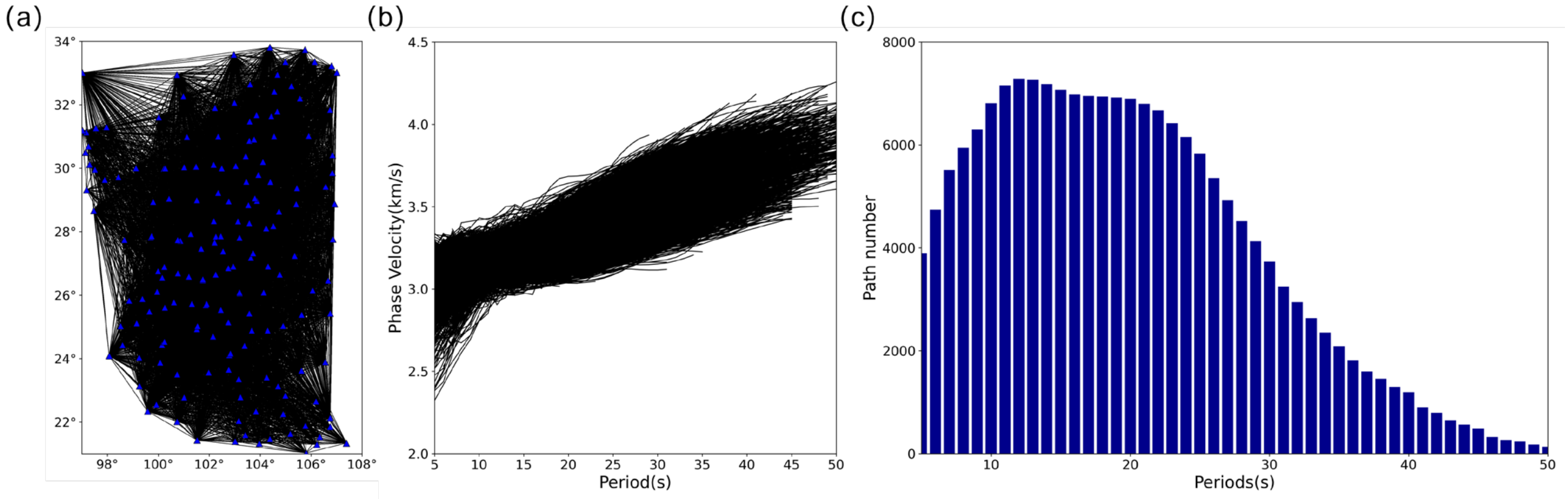

**Figure 7.** (a) The distribution of ray paths of surface-wave dispersion data. The blue triangles denote 224 stations used to extract interstation ambient noise Rayleigh-wave dispersion data. (b) Rayleigh-wave phase-velocity dispersion curves at periods from 5 s to 50 s. (c) Number of travel times (blue bars) at different periods.

We apply PSVI and sSVGD to infer the 3D shear-wave velocity (Vs) structure of the crust and upper mantle in southwest China, and compare their results. The velocity model is parameterized using a regular grid of 0.5° × 0.5° in the horizontal direction. In the vertical direction, the grid spacings are set to be 2 km from the sea level down to 10 km depth, 5 km from 10 to 50 km depth, 10 km from 50 to 70 km depth, and 30 km from 70 to 100 km depth to take account of decreasing sensitivity of surface waves when depth increases. Above the sea level, the grid spacings are set to be the altitude at each geographic point. We set data noise level to be 2.036 s according to previous studies (Liu et al. 2021). The prior distribution at each grid node is defined as an independent uniform distribution centered on a smoothed 3-D Vs model from Yang *et al.* (2020), with intervals increasing linearly from 1.0 km/s to 2.0 km/s as a function of depth.

Similar as above, PSVI is initialized with a standard Gaussian distribution $N(\theta|0, I)$ in the unconstrained parameter space, and a correlation kernel of size 5×5×5 is used to capture principal spatial dependencies (the cells in red box, Fig. 1). The distribution is then updated for 2000 iterations using 8 Monte Carlo samples per iteration, from which we draw 1000 samples and transformed them back to the original space to compute summary statistics. For sSVGD, we generated 40 particles from the prior distribution and updated them using the algorithm for 2000 iterations including a burn-in period of 1000 iteration. Similarly as above, we only retain every second sample after the burn-in period to save memory and storage cost. This results in a total of 20,000 samples, which are used to calculate statistics of the estimated posterior pdf.

**4.2 Results**

Fig. 8 shows horizontal slices of the mean and standard deviation at depths of 5, 10, 20, 30, 40, and 50 km obtained using sSVGD. At 5 km depth, the mean velocity mainly reflects shallow geological features. For example, the Sichuan Basin generally shows lower velocity than the rest of areas in Tibet. Within the Sichuan Basin, there are two low velocity anomalies in the northwest and south, and a relatively higher velocity anomaly in the center which is likely related to the uplift zone in this region due to extrusion of Longquan Shan reverse fault (Zhang *et al.* 2022). Note that the value of the southern low velocity anomaly is higher than that of the northwestern anomaly, which probably indicates thicker sediment accumulation in the northwest. At 10 km depth, there are low velocity anomalies around the XSHF, the LTF and the CHF, which probably reflect activities of those faults. The Sichuan Basin displays similar low velocity anomalies in the northwest and south and high velocity anomalies in the central as at the 5 km depth. In addition, there are high velocity anomalies around several reverse faults (LMSF, LJXJF) and inside the inner ELIP. At 20-30 km depths, there are two low-velocity zones (LVZs): one located west of the LMSF-LJXJF system and the other situated near the Xiaojiang Fault (XJF), which indicate the mechanically weak zones in the middle to lower crust, and has also been reported in a range of previous studies (Yang *et al.* 2020; Liu *et al.* 2023). These LVZs are separated by high velocity anomalies in the region around the LZJF and the ANHF, which corresponds to the inner and intermediate zones of the ELIP. In the Sichuan Basin, in contrast to the heterogeneous low-velocity anomalies observed at 10 km depth, the results exhibit a coherent high-velocity anomaly. At 40–50 km depths, the structure of the mean velocity model primarily reflects variation of Moho depths. In the east the region has transitioned into the upper mantle at this depth (Wang *et al.* 2017), displaying a relatively high velocity anomaly, whereas in the west the results show lower velocity because the lithology remains dominantly crustal. In addition, note that low velocity anomalies are observed across all depths beneath the Tengchong volcano (Shen *et al.* 2022).

Fig. 9 shows vertical profiles of the mean and standard deviation obtained using sSVGD at longitude of 31°N (AA'), 28°N (BB'), and 26°N (CC') respectively. The results reveal several LVZs between 10 and 30 km depth (regions contoured with 3.4 km/s), which have also been reported in a range of previous studies (Yang *et al.* 2020; Zhang *et al.* 2020; Liu *et al.* 2021, 2023; Qiao *et al.* 2018; Li *et al.* 2024). For example, in the northwest of the region the relatively large scale LVZ (LVZ1 in Fig. 8 and 9) can be clearly observed in profile AA′ and BB′. In AA′ the LVZ1 exhibits lateral discontinuity between the LTF and the XSHF, which is different from those found in previous studies (Yang *et al.* 2020; Liu *et al.* 2021, 2023) where the low velocity zone extends from the LTF to the XSHF. In profile BB′ the LVZ1 is thin in the west and becomes broader between the ZDF and the LXJF, similarly to previous findings (Liu *et al.* 2021, 2023). In profile CC', apart from the LVZ1 in the west, the relatively small-scale LVZ (LVZ2 in Fig. 8) can be observed beneath the XJF. In addition, the model reveals prominent high-velocity anomalies beneath the LZJF zone (profile CC') and the ANHF zone (profile BB'), which indicates the more rigid crust from the inner zone of the ELIP to the region around the ANHF zone, and likely obstructs the southeastward expansion of crustal material from the Tibetan Plateau (Bao *et al.* 2015; Qiao *et al.* 2018). At greater depths (> 40 km) the mean velocity model shows higher velocity, indicative of upper mantle composition, and the transition depth between the low and high velocity matches the Moho depth from Chen *et al.* (2021).

Overall the standard deviation models show increasing uncertainties with depth because of decreasing sensitivity of surface waves when depth increases (Fig. 8 and 9). In addition, there are localized lower uncertainties associated with velocity anomalies. For example, at shallow depths (< 20 km), the results show lower standard deviations at the location of higher velocity anomalies around the LMSF and LXJF (Fig. 8a and 8b), which is probably because rays concentrate in these anomalies. Similar phenomena can also be observed at other depths. Note that in vertical profiles the uncertainty is high at both edges because of lower data coverage as one would expect.

In Fig. 10 and 11, we show the results obtained using PSVI. Overall, the mean velocity model exhibits very similar structures to those obtained using sSVGD, including the low and high velocity anomalies in the Sichuan Basin at shallow depths (< 20 km), the low velocity zones in the mid-to-lower crust, and the velocity contrast between the western and eastern part at greater depths (> 20km). However, there are still small features that differ in the two results. For example, the shape and scale of the low velocity zones obtained using PSVI are different from those obtained using sSVGD (comparing Fig. 9 and Fig. 11). In addition, the standard deviation model obtained using PSVI show smoother structure and smaller magnitude than those from sSVGD, which probably reflects the limitation of PSVI as the method assumes Gaussian distribution with short-distance correlations. Similar observations have also been reported in earlier work (Ely *et al.* 2018; Zhang & Curtis 2020b; Wang *et al*. 2023; Zhao & Curtis 2024a). By contrast, sSVGD provides a more flexible approximation to the posterior distribution by using a set of samples, and therefore may produce more accurate results.

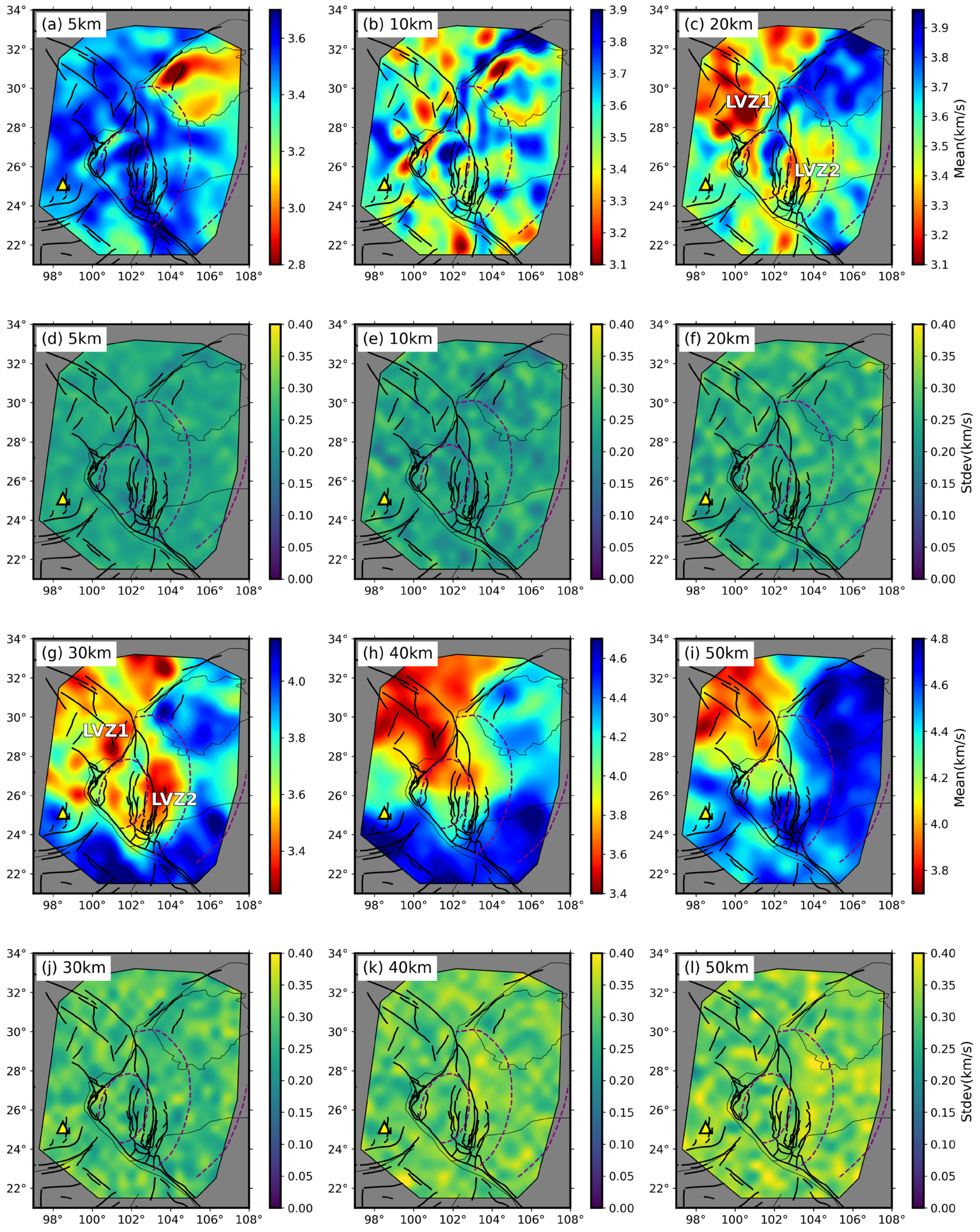

**Figure 8.** Horizontal slices of the mean and standard deviation of shear velocity obtained using sSVGD at depths of 5, 10, 20, 30, 40 and 50 km. The thin black lines outline the block boundaries, and the thick black lines represent major faults in this region. The purple dashed lines outline the boundaries between the inner, intermediate and outer zones of the ELIP. The yellow triangle denotes the location

of the Tengchong volcano.

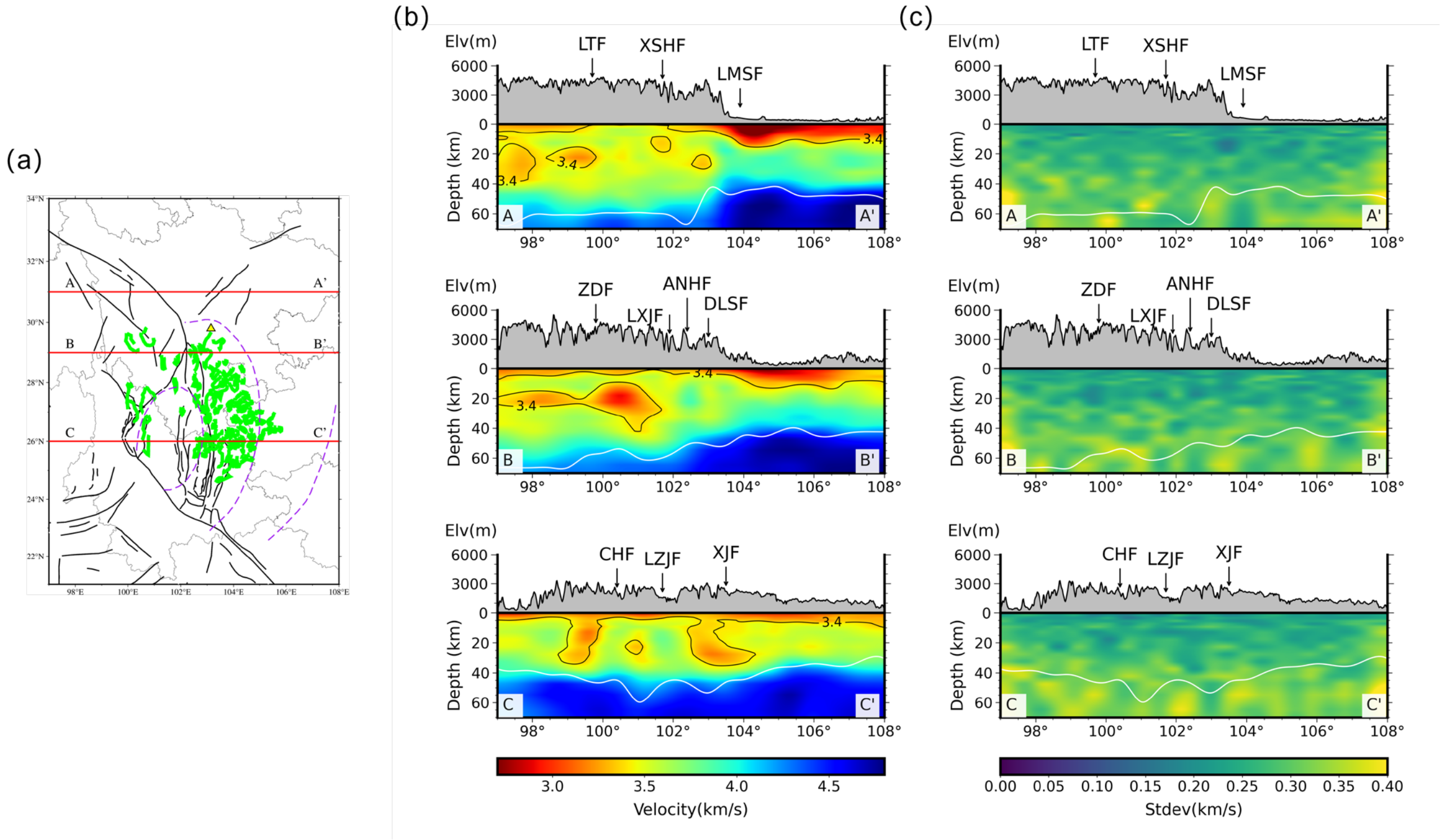

**Figure 9.** (a) Locations of the vertical profiles AA′, BB′ and CC′ along the longitude direction. Green shaded areas represent the distribution of basalts. Purple dashed lines denote the inner, intermediate and outer boundary of the ELIP. (b) The mean and (c) standard deviation of shear wave velocity along the three vertical profiles obtained using sSVGD. White solid lines show the Moho depth from Chen *et al.* (2021).

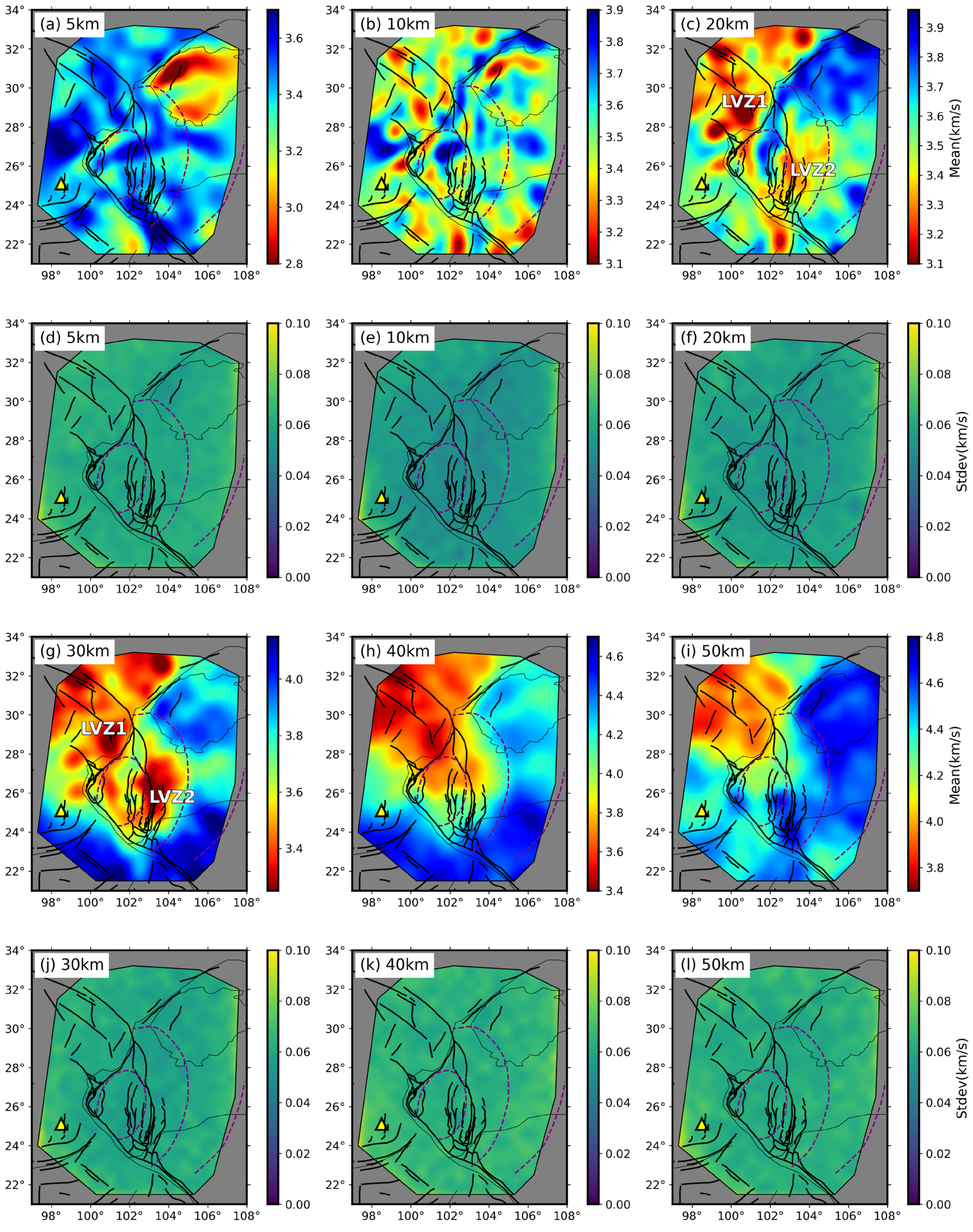

**Figure 10.** Horizontal slices of the mean and standard deviation obtained using PSVI. Keys as in Fig. 8.

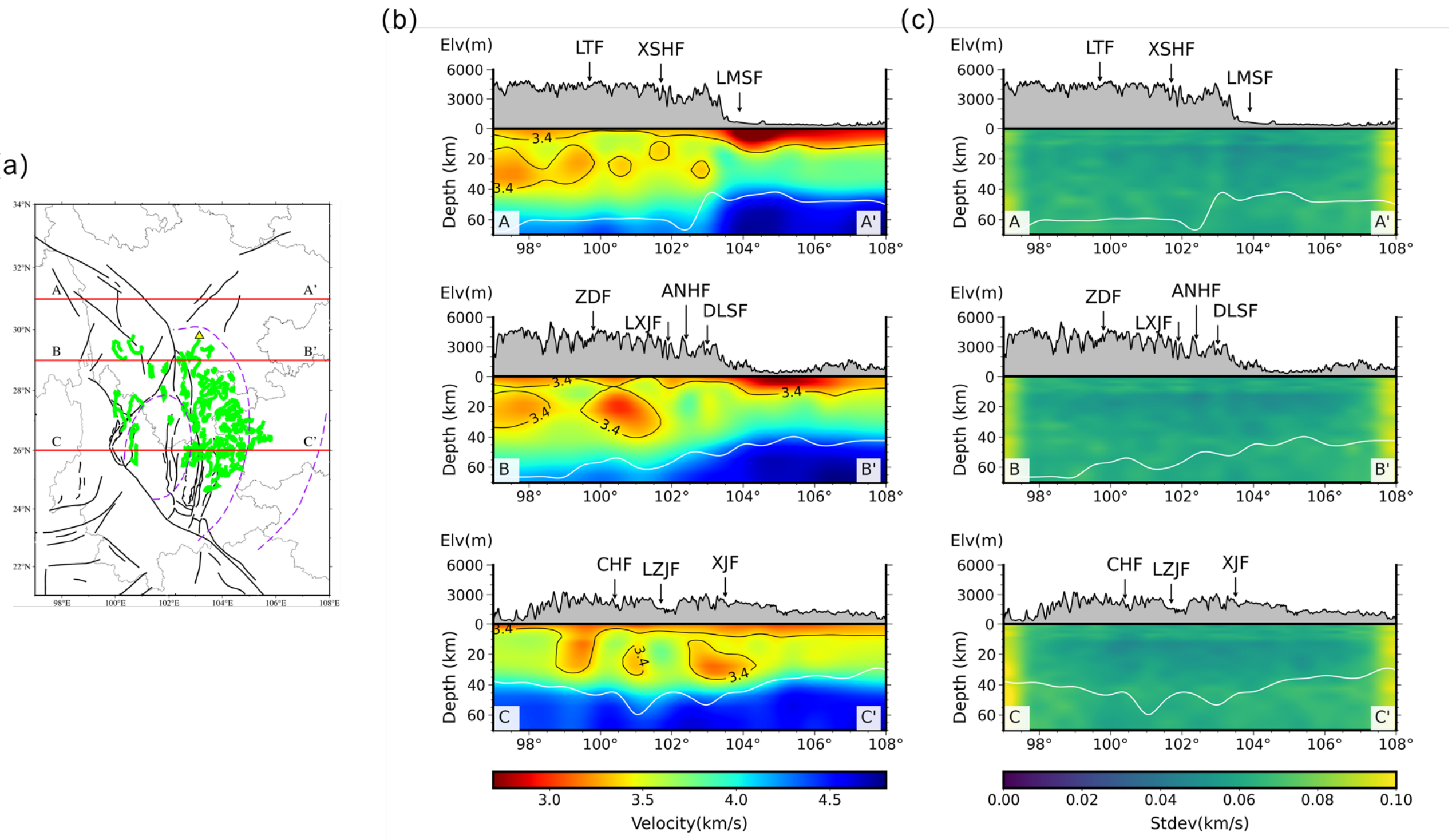

**Figure 11.** Vertical profiles of the mean and standard deviation obtained using PSVI. Kes as in Fig. 9.

### 4.3 Model comparison

We compare above models with two other velocity models obtained in previous studies using traditional methods: SWChinaCVM-1.0 (Liu *et al.* 2021) and SWChinaCVM-2.0 (Liu *et al.* 2023) from joint inversion of body and surface wave data (Fig. 12). Overall, those different models show similar features, for example, the low velocity anomalies in the northwest and south of the Sichuan basin at 10 km depth, the low velocity zones in middle-lower crustal, the high velocity anomaly within the inner zone of the ELIP at 20km depth, and the velocity contrast between the eastern and western region at 40 km depth. However, there still exists details that differ in those models. For example, the models obtained using sSVGD and PSVI show many low velocity anomalies around faults in regions outside the Sichuan Basian at 10 km depth, which cannot be clearly observed in those obtained using traditional methods. This discrepancy may arise from strong regularization constraints applied in the traditional inversions, which can suppress small-scale heterogeneities (Zhadov 2002; Yang *et al.* 2025). Within the Sichuan Basin, the models obtained using sSVGD and PSVI show stronger low velocity anomalies in the northwest and south compared to the SWChinaCVM-1.0 model, and are more similar to the SWChinaCVM-2.0 model which is obtained using more data (Liu *et al.* 2023). This demonstrates that variational methods can probably obtain more accurate results than traditional methods. Similar phenomena can also be observed at 20 and 40 km depth. For example, the models obtained using sSVGD and PSVI show more detailed low velocity structures within the ELIP than those in the SWChinaCVM-1.0 model, which is also more similar to the SWChinaCVM-2.0 model. However, we note that these detailed features do not necessarily reflect real structures of the Earth, because these are the mean structures of a set of possible models, and also because the algorithms may not have converged sufficiently in such a high-dimensional problem. Nevertheless, the similarity between different models obtained using different methods provides confidence for these structures.

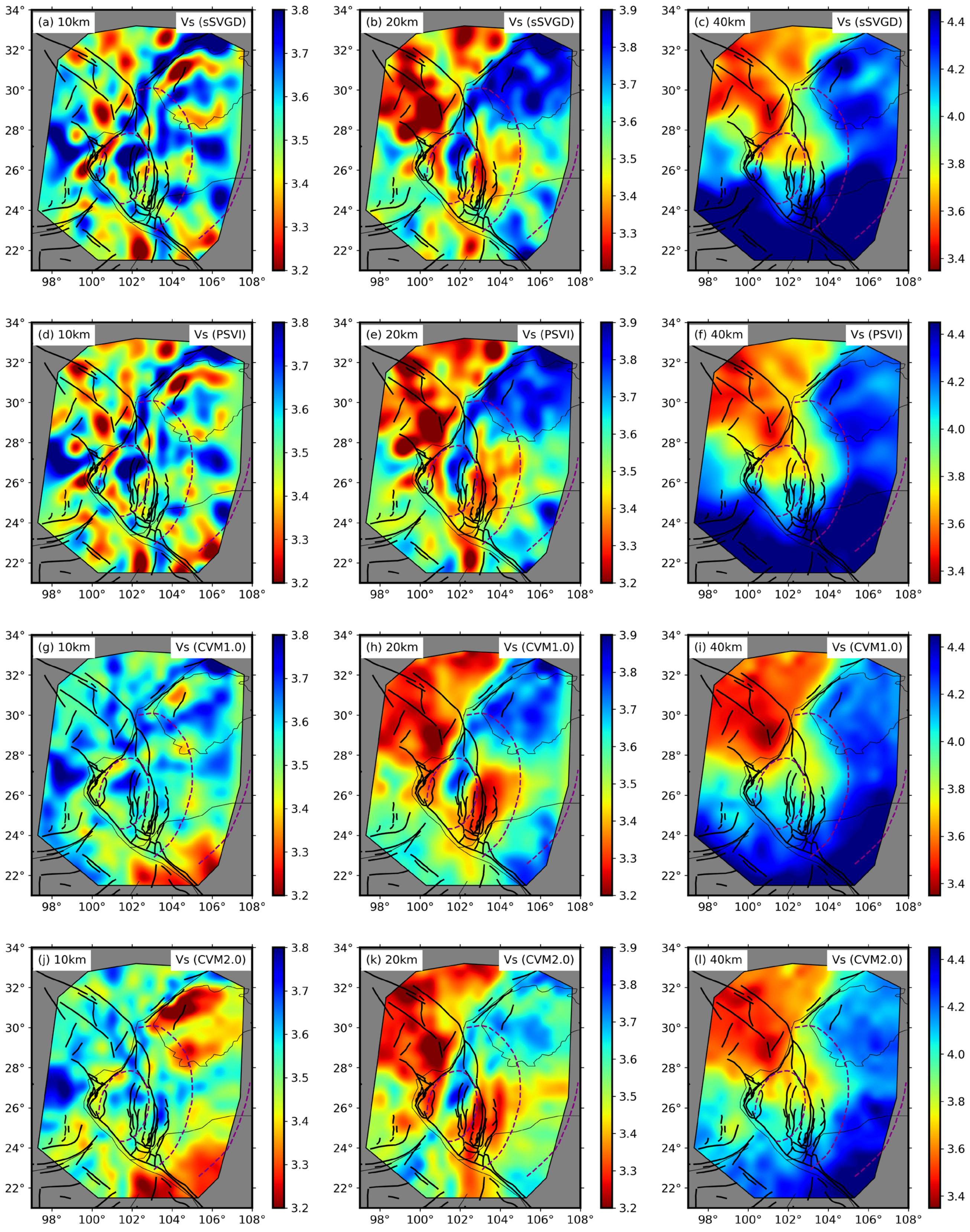

**Figure 12**. Horizontal slices of different *V*s models at 10, 20 and 40 km depth. (a)–(c) *V*s model from sSVGD. (d)–(f) *V*s model from PSVI. (g)–(i) SWChinaCVM-1.0. (j)–(l) SWChinaCVM-2.0. Keys as in Figure 8.

## 5. Discussion

We demonstrated that variational inference methods can be effectively applied to 3D surface wave tomography. The mean-field ADVI method underestimates posterior uncertainties because the method neglects correlation information between parameters. PSVI improves the results accuracy by exploiting a sparse-structured covariance matrix which only includes correlation between parameters at proximal cells, while maintaining a computational cost comparable to mean-field ADVI. By contrast, sSVGD requires more computational cost than mean-field ADVI or PSVI because the method uses a set of particles to approximate the posterior distribution. However, sSVGD can generate more accurate results than mean-field ADVI or PSVI as the method can theoretically approximate the posterior distribution to arbitrary accuracy. To further improve efficiency of sSVGD, high order gradient information may be used, for example, using a Hessian matrix kernel (Wang *et al.* 2019) or the stochastic Stein variational Newton method (Leviyev *et al.* 2022). In addition, one may try to reduce dimensionality of the problem to reduce the number of required samples. For example, one may use other parameterizations that require fewer parameters to represent the velocity model, such as Voronoi cells (Bodin & Sambridge 2009; Zhang *et al.* 2018), wavelet parameterization (Hawkins & Sambridge 2015), discrete cosine transforms (Urozayev *et al.* 2022), or neural network reparameterization (Laloy *et al.* 2017; Mosser *et al.* 2020; Agata *et al.* 2023).

For the likelihood function we used a Gaussian distribution with a fixed data noise level in the variational methods. In practice this noise level should be determined from the data. For example, one can use the standard deviation of daily measurements as the noise level, or estimate it using the maximum likelihood method (Sambridge 2014). It may also be possible to estimate the noise level in the inversion process using a hierarchical Bayesian formulation (Malinverno & Briggs 2004; Ranganath *et al.* 2016). In addition, other non-Gaussian likelihood functions may be used to improve the results given that those likelihood functions represent the distribution of data uncertainty (Zhang *et al.* 2023).

In this study we used a smoothness constrained prior distribution, and assessed the influence of such constraints on the posterior results. However, in many practical applications it is challenging to define an appropriate prior probability distribution over smoothness levels. To reduce this issue, hierarchical Bayesian methods may be used to estimate this distribution a posteriori, but such approaches can introduce physical inconsistencies into Bayesian inversion outcomes (Mosegaard & Curtis 2024). In practice where more knowledge about the subsurface is available, one can use a more informative prior distribution, for example, by incorporating information such as geological classifications, well logs, subsurface structures and expert knowledge into the prior distribution. In addition, other forms of prior distribution may be used, for example, Gaussian process (Ambikasaran *et al.* 2015; Liu *et al.* 2020), or neural network-based prior distribution that encode geological features into the prior distribution (Laloy *et al.* 2017; Mosser *et al.* 2020).

For computational efficiency we used a forward modelling method based on modal approximation and 2-D fast marching method. To further improve accuracy, full waveform modelling methods may be used to simulate surface wave data (Tromp *et al.* 2005; Fichtner *et al.* 2006). However, we note that this requires expensive computational cost (Zhang *et al.* 2023). In this study we only used surface wave dispersion data to constrain the subsurface. In future, other types of data, for example, body wave travel time data and receiver functions, can also be included to improve the results (Julia *et al.* 2000; Fang *et al.* 2016; Zhang *et al.* 2020).

## 6. Conclusions

In this study we applied a set of variational inference methods (mean-field ADVI, PSVI and sSVGD) to solve 3-D surface wave tomographic problems using both synthetic and real data. The results show that mean-field ADVI can provide rapid solutions but with systematically underestimated uncertainty. PSVI improves the results accuracy by incorporating structured correlations among parameters, while maintaining computational efficiency of mean-field ADVI. sSVGD appears to be the most expensive method, but provides the most accurate results with reasonable uncertainty estimates. In real data applications from Southwest China, the variational methods provide more detailed structures than those obtained using traditional methods, along with reliable uncertainty estimates. We therefore conclude that variational inference methods can be applied fruitfully to address real-world 3D surface wave tomographic problems.

**Appendix A: Construction of the correlation matrix L in PSVI**

In this section, we present a method to construct the three-dimensional correlation matrix **L** given a specific correlation length.

Assume a correlation length of two cells in the $x$-direction we can construct a lower-triangular sparse matrix with two sub-diagonals:

$$D_x = \begin{bmatrix} l_{0,1} & & & & & & \\ l_{1,1} & l_{0,2} & & & & & \\ l_{2,1} & l_{1,2} & l_{0,3} & & & & \\ & l_{2,2} & l_{1,3} & \ddots & & & \\ & & l_{2,3} & \ddots & \ddots & & \\ & & & \ddots & \ddots & l_{0,nx-2} & & \\ & & & & \ddots & l_{1,nx-2} & l_{0,nx-1} & \\ & & & & & l_{2,nx-2} & l_{1,nx-1} & l_{0,nx} \end{bmatrix}$$

Similarly, we can obtain the D matrices in the $y$ direction $(D_y)$ and the $z$ direction $(D_z)$. The number of off-diagonal blocks in the D matrix can be controlled by the number of adjacent grid points considered in each spatial direction.

Given the above definition, the correlation matrix for parameters on a regular 3-D grid can be obtained by multiplying the $D_x, D_y$ $and$ $D_z$ in sequence using the Kronecker product:

$$\boldsymbol{L} = D_x \otimes (D_y \otimes D_z)$$

where $\otimes$ represents the Kronecker product which is defined as:

$$D_y \otimes D_z = \begin{bmatrix} d_{1,1}D_z & \cdots & d_{1,ny}D_z \\ \vdots & \ddots & \vdots \\ d_{ny,1}D_z & \cdots & d_{ny,ny}D_z \end{bmatrix}$$

where $(d_{1,1}, \dots, d_{ny,ny})$ represents elements of $D_y$, and $ny$ is the number of grid points in $y$ direction. Note that the order of the matrix multiplication must correspond to the order of parameter expansion as the Kronecker product is non-commutative.

**Appendix B：Synthetic tests for different prior hypotheses**

In this section, we compare a smoothed prior distribution $(\sigma_x = \sigma_y = 0.2, \sigma_z = 1.0)$ with a non-smoothed uniform prior distribution to demonstrate the effect of smoothness constraints. The inversions are performed in exactly the same way for different prior distributions.

Fig. B1 presents the results for a horizontal section at a depth of 20 km using both prior pdfs (smoothed and non-smoothed). The first two rows present the mean and standard deviation obtained without smoothness, while the last two rows show those with smoothness applied. From left to right, each column corresponds to the results obtained using mean-field ADVI, PSVI, and sSVGD, respectively.

Fig. B2 compares the posterior marginal pdfs at two vertical profiles (locations are indicated with white pluses in Fig. B1) obtained using non-smoothed and smoothed prior distributions with different methods. The results show broader distributions when using smoothed prior pdfs as one would expect. Note that in the case with non-smoothed prior pdfs, the results obtained using sSVGD show more complex distributions than those obtained using mean-field ADVI and PSVI, which may reflect limitation of mean-field ADVI and PSVI as they assume Gaussian assumptions.

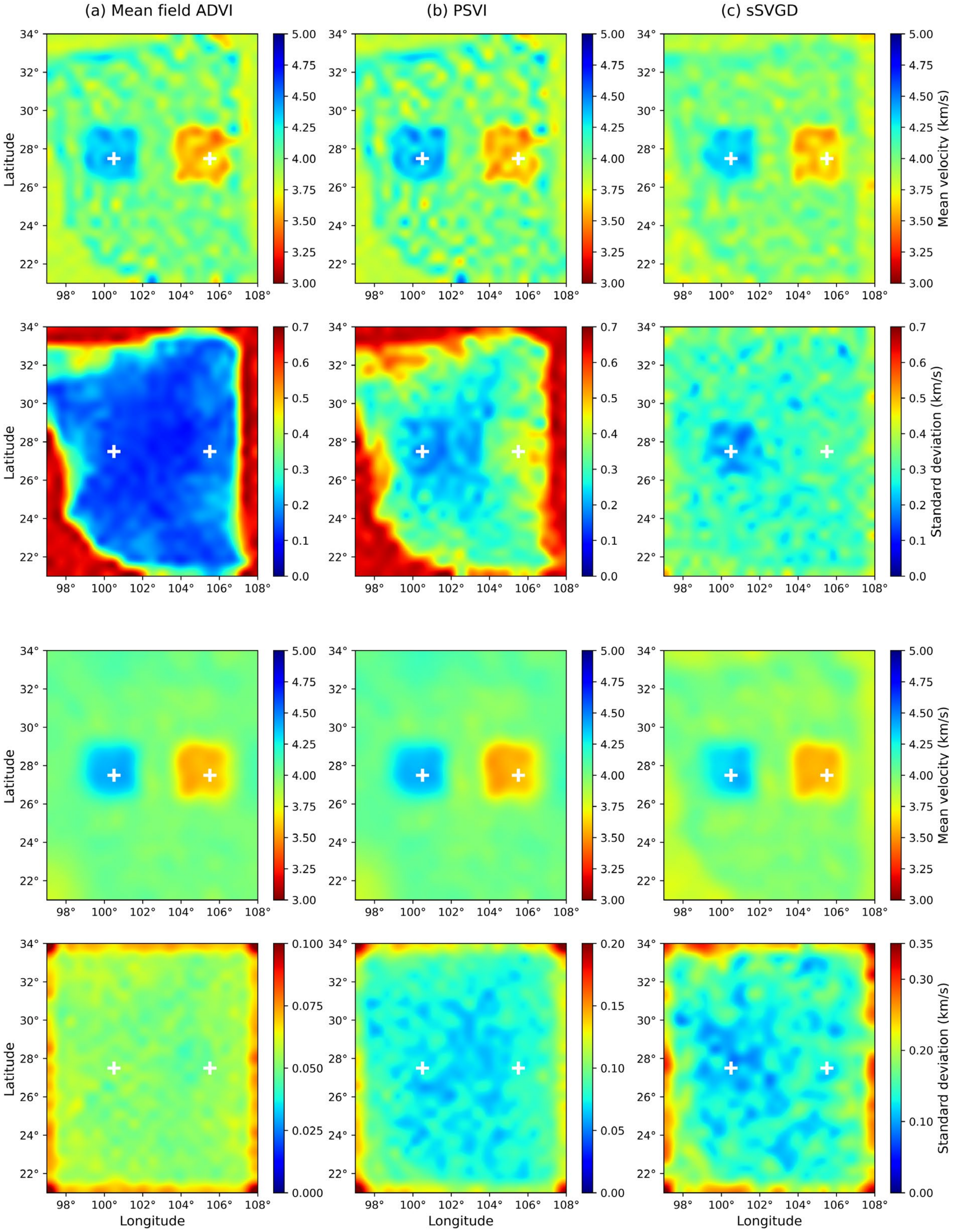

**Figure B1.** Inversion results on a horizontal section at 20 km depth obtained using (a) Mean Field ADVI, (b) PSVI, and (c) sSVGD. The top panels show the results without smoothness, and the lower panels show those with smoothness applied ($\sigma_x = \sigma_y = 0.2, \sigma_z = 1.0$).

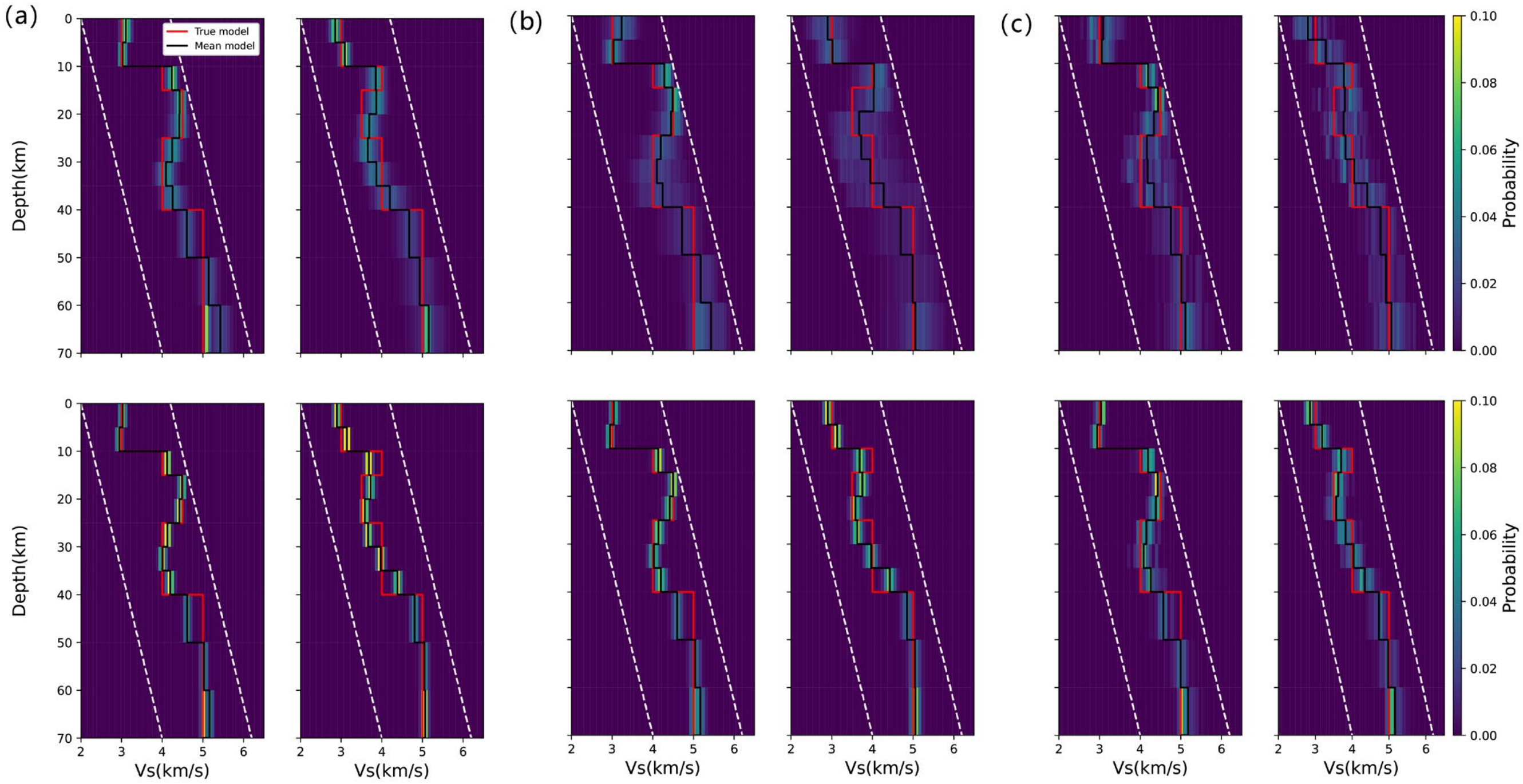

**Figure B2.** Posterior marginal probability density functions (pdfs) at two vertical locations, as indicated by the white pluses in Figure B1. (a) Results from mean-field ADVI. The top panels show the results obtained using the non-smoothed uniform prior pdf, and the low panels show those obtained using the smoothed prior pdf with parameters $\sigma_x = \sigma_y = 0.2$, and $\sigma_z = 1.0$. (b) and (c) present the posterior marginal pdfs obtained using PSVI and sSVGD, respectively.

## Acknowledgments

This study is supported by the National Natural Science Foundation of China (Grant Nos. 42204055 and U23B20160), and the Fundamental Research Funds for the Central Universities of China University of Geosciences in Beijing. This work used the HPC service of China University of Geosciences Beijing (https://hpc.cugb.edu.cn).